\documentclass{jfm}
\usepackage{graphicx}
\usepackage{epstopdf, epsfig}
\usepackage{amsmath}
\usepackage{bm}
\usepackage{color}

\shorttitle{Thermal convection with an inclined magnetic field}
\shortauthor{S. Bhattacharya, T. Boeck, D. Krasnov and J. Schumacher}

\title{Wall-attached convection under strong inclined magnetic fields}

\author{Shashwat Bhattacharya\aff{1}
  \corresp{\email{shashwat.bhattacharya@tu-ilmenau.de}},
  Thomas Boeck\aff{1},
  Dmitry Krasnov\aff{1}
 \and J{\"o}rg  Schumacher\aff{1,2}}

\affiliation{\aff{1}Institute of Thermodynamics and Fluid Mechanics, Technische Universit{\"a}t Ilmenau,
Postfach 100565, D-98684 Ilmenau, Germany
\aff{2}Tandon School of Engineering, New York University, New York 11021, USA}

\begin{document}

\maketitle

\begin{abstract}
We employ a linear stability analysis and direct numerical simulations to study the characteristics of wall-modes in thermal convection in a rectangular box under strong and inclined magnetic fields. The walls of the convection cell are electrically insulated.  The stability analysis assumes periodicity in the spanwise direction perpendicular to the plane of the homogeneous magnetic field. Our study shows that for a fixed vertical magnetic field, the imposition of horizontal magnetic fields results in an increase of the critical Rayleigh number along with a decrease in the wavelength of the wall modes. The wall modes become tilted along the direction of the resulting magnetic fields and therefore extend further into the bulk as the horizontal magnetic field is increased. Once the modes localized on the opposite walls interact, the critical Rayleigh number decreases again and eventually drops below the value for onset with a purely vertical field. 
We find that for sufficiently strong horizontal magnetic fields, the steady wall modes occupy the entire bulk and therefore convection is no longer restricted to the sidewalls. 
The above results are confirmed by direct numerical simulations of the nonlinear evolution of magnetoconvection. 
\end{abstract}

\section{Introduction}\label{sec:Introduction}
Buoyancy-driven flows of electrically conducting fluids under the influence of magnetic fields are a common occurrence in geophysical, astrophysical, as well as in several technological applications. Such flows are called magnetoconvection and their driving mechanism is the temperature dependence of the fluid density, which results in spatial density variations leading to buoyancy forces acting on the fluid. When such a fluid moves under the influence of magnetic fields, electric currents are induced in the fluid due to Faraday's law, which, in turn, induce magnetic fields by the virtue of Ampere's law. These electric currents interact with the applied and induced magnetic fields to generate a Lorentz force distribution that acts on the fluid. Therefore, magnetoconvective flows are governed by equations of conservation of mass, momentum, and thermal energy, along with Maxwell's equations for electromagnetism and Ohm's law~\citep{Weiss:book}.
Magnetoconvection is encountered in the Sun, stars, and planetary dynamos. In industries and technological applications, magnetoconvection is typically encountered in liquid-metal batteries~\citep{Kelley:PF2014, Shen:TCFD2016,Kelley:AMR2018}, cooling liquid-metal blankets in fusion reactors~\citep{Mistrangelo:FED2020,Mistrangelo:FED2021}, and magnetic stirring and braking of liquid metal melts~\citep{Davidson:ARFM1999,Davidson:book:MHD,Lyubimov:JFM2010}.

A simplified paradigm for magnetoconvection consists of a fluid layer that is heated from below and cooled from above (Rayleigh-B{\'e}nard convection or RBC) with imposed magnetic fields in different configurations.
Typically, the Boussinesq approximation is employed for modelling the above flows; this approximation assumes that the flow is incompressible and the density variations are negligible except in the buoyancy term in the momentum equation~\citep{Chandrasekhar:book:Instability,Lohse:ARFM2010,Chilla:EPJE2012,Verma:book:BDF}. 
Magnetoconvection is governed by the following nondimensional parameters: i) Rayleigh number $\Ray$ -- the ratio of buoyancy to dissipative forces, ii) Prandtl number $\Pran$ -- the ratio of kinematic viscosity to thermal diffusivity, iii) Hartmann number $\Ha$ -- the ratio of Lorentz to viscous forces, and iv) the magnetic Prandtl number $\Pm$ -- the ratio of kinematic viscosity to magnetic diffusivity.
The important nondimensional output parameters of magnetoconvection are i) the Nusselt number $\Nu$ -- the ratio of the total heat transport to the diffusive heat transport, ii) the Reynolds number $\Rey$ -- the ratio of inertial to viscous forces, and iii) the magnetic Reynolds number $\Rm$ -- the  ratio of induction to diffusion of the magnetic field.
In liquid-metal convection typically encountered in laboratory experiments and most industrial applications, the magnetic Reynolds number is sufficiently small such that the induced magnetic field is negligible compared to the applied magnetic field and is thus neglected in the expressions of the Lorentz force and Ohm's law~\citep{Roberts:book,Davidson:book:MHD,Verma:ET}. 
Such cases are referred to as {\em quasi-static} magnetoconvection where the induced magnetic field adjusts instantaneously to the changes in velocity. 
In the quasi-static approximation, there exists a one-way influence of the magnetic field on the flow only.

Magnetoconvection has been studied theoretically in the past~\citep[for example,][]{Chandrasekhar:book:Instability,Houchens:JFM2002,Busse:PF2008} as well as with the help of experiments~\citep[for example,][]{Nakagawa:PRSL1957,Fauve:JPL1981,Cioni:PRL2000,Aurnou:JFM2001,Burr:PF2001,King:PNAS2015,Vogt:PRF2018,Vogt:JFM2021,Zuerner:JFM2020,Grannan:JFM2022} and numerical simulations~\citep[for example,][]{Liu:JFM2018,Yan:JFM2019,Akhmedagaev:MHD2020,Akhmedagaev:JFM2020,Nicoski:PRF2022,Bhattacharya:JFM2023}.
An application of horizontal magnetic fields causes the large-scale rolls to become quasi two-dimensional and align in the direction of the field~\citep{Fauve:JPL1981,Busse:JTAM1983,Burr:JFM2002,Yanagisawa:PRE2013,Tasaka:PRE2016,Vogt:PRF2018,Vogt:JFM2021}. 
These self-organized flow structures reach an optimal state wherein the heat transport and convective velocities increase significantly compared to convection without magnetic fields~\citep{Vogt:JFM2021}. 
In contrast, strong vertical magnetic fields suppress convection~\citep{Chandrasekhar:book:Instability,Cioni:PRL2000,Zuerner:JFM2020,Akhmedagaev:MHD2020,Akhmedagaev:JFM2020}. 
It must be noted that in a Rayleigh-B{\'e}nard system, convection commences only above a certain critical Rayleigh number, which is  $\Ray_c\approx 1708$ for the case with infinite no-slip horizontal walls~\citep{Chandrasekhar:book:Instability}.
For $\Ray<\Ray_c$, the heat transfer occurs purely by diffusion.
The critical Rayleigh number increases when a vertical magnetic field is imposed and scales as $\Ray_c \sim \Ha^2$ in the asymptotic limit of large Hartmann numbers.

The dynamics of convection under strong vertical magnetic fields become more intricate in the presence of sidewalls. \citet{Houchens:JFM2002} and \citet{Busse:PF2008} analytically showed that magnetoconvection near the sidewalls ceases at Rayleigh numbers below the ones required to completely suppress convection in the bulk. Several numerical and experimental studies on magnetoconvection with sidewalls have also revealed the presence of residual wall-attached convection at $\Ray<\Ray_c$~\citep{Houchens:JFM2002,Liu:JFM2018,Akhmedagaev:MHD2020,Akhmedagaev:JFM2020,Zuerner:JFM2020,teimurazov2023unifying,mccormack2023wall}. These so-called wall modes were shown to exhibit a two-layered structure and become more closely attached to the sidewalls with the increase of Hartmann number~\citep{Liu:JFM2018}. 

There are a few studies only on convection with inclined magnetic fields which motivates the present work~\citep{HurlBurt:APJ1996,Nicoski:PRF2022}. The results of \citet{HurlBurt:APJ1996} indicate that the mean flows tend to travel in the direction of the tilt. \citet{Nicoski:PRF2022}  observed qualitative similarities between convection with inclined magnetic field and that with vertical magnetic field in terms of the behavior of convection patterns, heat transport, and flow speed. However, to the best of our knowledge, there are no studies for the case with inclined magnetic fields where the Rayleigh number is close to but less than the critical Rayleigh number. Therefore, in the present work, we study thermal magnetoconvection in the wall-attached convection regime and explore the effects of additional horizontal magnetic fields on the wall modes. We use a combination of linear stability analysis and direct numerical simulations to study the dependence of the horizontal magnetic field strength, relative to the vertical magnetic field, on the wall-mode structures and their impact on large-scale heat and momentum transport.

The outline of the paper is as follows. In \S~\ref{sec:Model}, we discuss the problem setup, linear stability model, and the schemes for direct numerical simulations. The linear stability analysis and the results of direct numerical simulations are described in \S~\ref{sec:Results}. We conclude in \S~\ref{sec:Conclusions}.   

\section{Numerical model}\label{sec:Model}
In this section, we discuss the mathematical model of our problem and the numerics employed for the stability analysis and direct numerical simulations.
The study will be conducted under the quasi-static approximation, in which the induced magnetic field is neglected as it is very small compared to the applied magnetic field. This approximation is fairly accurate for magnetoconvection in liquid metals~\citep{Davidson:book:MHD}.
Further, we employ Boussinesq approximation, in which the variations in the density of the fluid are ignored except in the buoyancy term in the momentum equation.
Hence, the flow is essentially treated as incompressible.
The governing equations of magnetoconvection under the above approximations are as follows:
\begin{eqnarray}
\nabla \cdot \boldsymbol{u}&=&0 \label{eq:continuity} \\
\frac{\partial \boldsymbol{u}}{\partial t}  + \boldsymbol{u}\cdot \nabla \boldsymbol{u} &=& -\frac{\nabla p}{\rho} + \alpha g T\hat{z}+ \nu\nabla^2 \boldsymbol{u}+ \frac{1}{\rho}(\boldsymbol{j} \times {\boldsymbol{B}}),
\label{eq:Momentum} \\
\frac{\partial T}{\partial t} + \boldsymbol{u} \cdot \nabla T &=& \kappa \nabla^2 T, \label{eq:T_energy} \\
\boldsymbol{j} &=& \sigma \left(-\nabla \phi + \boldsymbol{u} \times {\boldsymbol{B}}\right),
\label{eq:Current}\\
\nabla^2 \phi &=& \nabla \cdot (\boldsymbol{u} \times {\boldsymbol{B}}),
\label{eq:Potential}
\end{eqnarray}
where $\boldsymbol{u}$, $\boldsymbol{j}$, $p$, $T$, and $\phi$ are the fields of velocity, current density, pressure, temperature, and electrical potential respectively, and  ${\boldsymbol{B}}=(B_x,B_y,B_z)$ is the applied magnetic field. 
In our work, the magnetic fields are inclined along $y$-direction only, hence $B_x=0$.
Further, $\nu$ is the kinematic viscosity, $\kappa$ is the thermal diffusivity, $\rho$ is the density, and $\sigma$ is the electrical conductivity of the fluid.
The last term in the momentum equation (\ref{eq:Momentum}) is the Lorentz force density. Equation (\ref{eq:Current}) is Ohm's law. The Poisson equation (\ref{eq:Potential}) for the electric potential  is a consequence of the charge conservation condition $\nabla\cdot \boldsymbol{j}=0$. 

In the following, we will discuss how the above equations have been employed for our linear stability analysis and direct numerical simulations.

\subsection{Linear stability model}
\label{sec:Linstab}
We first discuss the derivation of the perturbation equations for our linear stability analysis. The equations~(\ref{eq:continuity}) to (\ref{eq:Potential}) are non-dimensionalized using the cell height $H$ as the length scale, $\kappa/H$ as the velocity scale, the temperature difference $\Delta$ between the two horizontal plates as the temperature scale, and $B_z$, the vertical component of the applied magnetic field. For this part of the analysis, we take the units that are typically chosen for a linear stability analysis to end with a Prandtl number-independent set of equations at the marginal stability threshold. The characteristic units in the subsequent simulation part will differ.
The non-dimensionalized governing equations are as follows.
 \begin{eqnarray}
\label{eq:momentum}
&&\frac{1}{Pr}\left(\frac{\partial \bm{u}}{\partial t} + (\bm{u} {\cdot \nabla})\bm{u}\right) =
- {\nabla p} +  {\nabla}^2 \bm{u} + \Ray \, \theta \bm{e}_z
+ \Ha_z^2 \,\bm{j}\times\left( \bm{e}_z+R \bm{e}_y\right),\\
\label{eq:energy}
&&\frac{\partial \theta}{\partial t} + (\bm{u} {\cdot \nabla})\theta =
  {\nabla}^2 \theta +u_z,\\
\label{eq:mass}
&&\nabla \cdot \bm{u}= 0,\\
\label{eq:Ohm}
&&\bm{j}=-\nabla \phi  + \bm{u} \times \left(\bm{e}_z+ R \bm{e}_y\right),\\
\label{eq:Ampere}
&&\nabla \cdot \bm{j} =0. 
\end{eqnarray}
In the above system of equations, $\Ray$ is the Rayleigh number, $\Pran$ is the Prandtl number, $\Ha_z$ is the Hartmann number based on the vertical component of the magnetic field, and $R$ is the ratio of the horizontal to the vertical magnetic field strength. These quantities are the governing parameters for our setup and are given by
\begin{equation}
\Ray = \frac{\alpha g \Delta H^3}{\nu \kappa}, \quad \Pran= \frac{\nu}{\kappa}, \quad \Ha_z = B_z H\sqrt{\frac{\sigma}{\rho \nu}}, \quad R = \frac{B_y}{B_z}.
\end{equation}
The quantity $\theta$ is the difference between the temperature and the  linear conduction profile, i.e.,
\begin{equation}
T({\bm x},t)=\theta({\bm x},t) -z\,.
\end{equation}
Apart from $\Ray$, $\Pran$, $\Ha_z$, and $R$, the dynamics are also governed by the aspect ratio $\Gamma$, which is the ratio of the length to the height of the convection cell. 

\begin{figure}
  \centerline{\includegraphics[scale = 0.2]{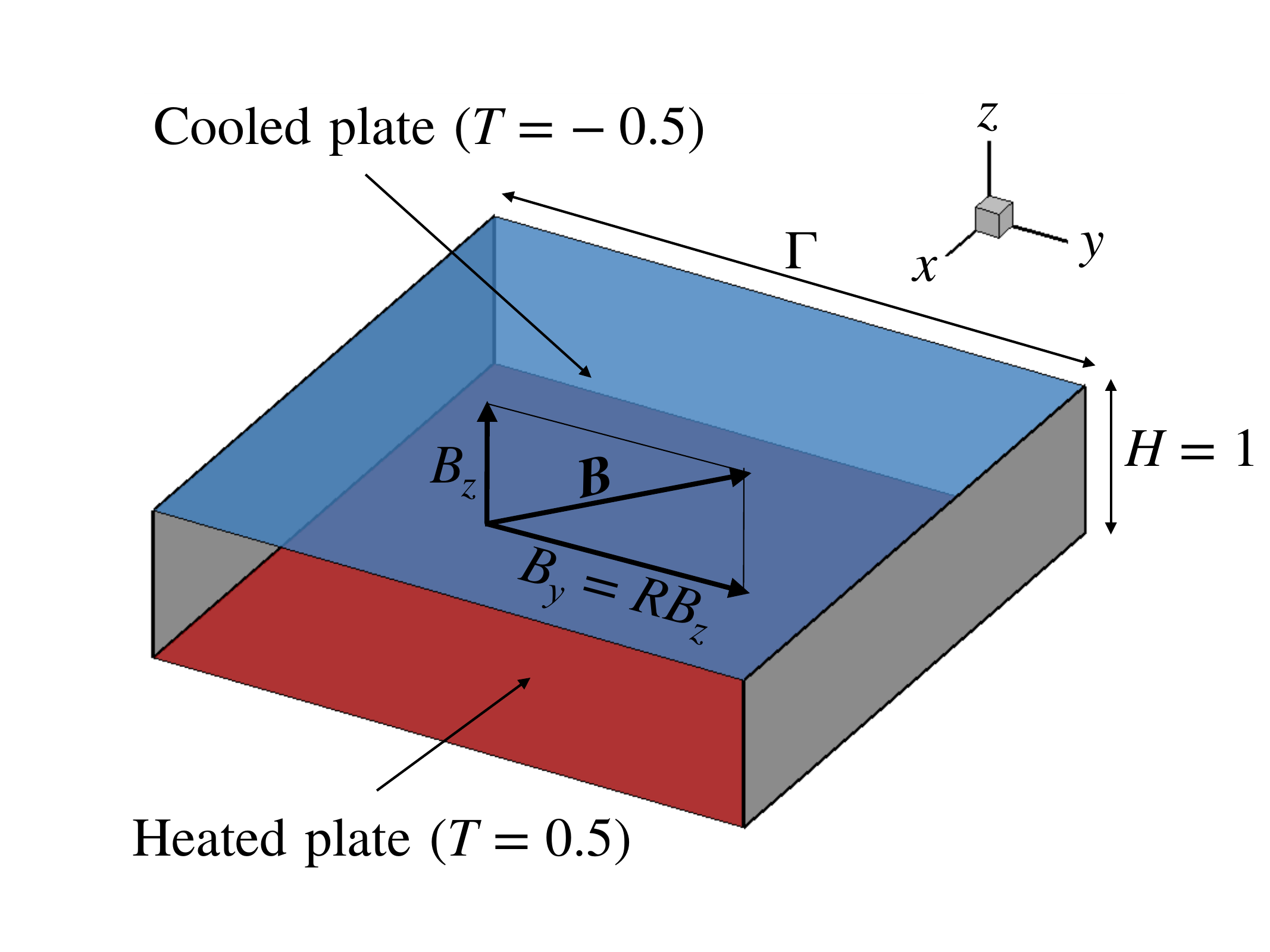}}
  \caption{A sketch of the Rayleigh-B{\'e}nard convection setup with inclined magnetic field employed for our linear stability analysis.
 }
\label{fig:Sketch}
\end{figure}

We examine a total of 12 cases of magnetoconvection in a bounded, horizontally-extended domain of dimension $\Gamma \times 1$ in the $y$-$z$ plane. 
Two aspect ratios are considered: $\Gamma=2$ and $\Gamma=4$. For $\Gamma=2$, we consider the cases of $\Ha_z=50$ and $\Ha_z=100$, whereas for $\Gamma=4$, we consider only the case of $\Ha_z=50$. For each $\Ha_z$ analysed in our study, we vary $R$ from 0 (corresponding to a purely vertical magnetic field) to 3 in steps of 1. The convection cell is periodic in $x$-direction and consists of no-slip horizontal walls at $z = \pm1/2$ and two no-slip sidewalls at $y= \pm \Gamma/2$. All the walls are electrically insulated. Each horizontal wall is at a constant temperature, with the bottom wall at $T=0.5$ and the top wall at $T=-0.5$. 
The sidewalls are thermally insulated with $\partial T/\partial \eta=0$, where $\eta$ is the direction normal to the wall. A sketch of our setup is shown in figure~\ref{fig:Sketch}.

In the present work we assume that the instability is of stationary type. The nonlinear terms as well as the time derivatives in
equations (\ref{eq:momentum}) and (\ref{eq:energy}) are therefore neglected. The momentum and continuity equations reduce to the Stokes problem with additional buoyancy and
Lorentz force. To avoid complications stemming from the coupling between pressure and velocity we choose a representation
for the velocity that satisfies the continuity equation automatically and eliminate the pressure term.  The velocity field is written as the curl of a  {\em vector streamfunction} $\bm{\psi}$  
 \begin{equation}
\label{eq:vectorstreamfunction}
\bm{u}=\nabla \times \bm{\psi},
\end{equation}
and the gauge condition  $\nabla\cdot \bm{\psi}=0$ is imposed to determine $\bm{\psi}$ uniquely as in \cite{Priede:JFM2010}. The dependence on  $x$ is represented by the normal mode ansatz with wavenumber $\beta$ for all fields, e.g. $\theta(y,z)\exp(i\beta x)$ for the temperature perturbation. The gauge condition  allows one to express the $x$-component of $\bm{\psi}$  by
\begin{equation}
\label{eq:psiy}
i \beta \psi_x (y,z)= -\partial_y \psi_y(y,z)-\partial_z \psi_z(y,z).
\end{equation}
The velocity components then read
\begin{align}
\label{eq:velocitycomponents1}
u_x &= \partial_y \psi_z -\partial_z \psi_y\,,\\
\label{eq:velocitycomponents2}
u_y &= -i\beta \psi_z +\frac{i}{\beta}\left(\partial_y \partial_z\psi_y+\partial_z^2 \psi_z\right)\,,\\
\label{eq:velocitycomponents3}
u_z &= i\beta \psi_y -\frac{i}{\beta}\left(\partial_y^2 \psi_y+\partial_y \partial_z \psi_z\right)\,.
\end{align}
Equations for $\psi_y$ and $\psi_z$ are obtained by taking the curl of the definition (\ref{eq:vectorstreamfunction}) and the momentum equation (\ref{eq:momentum}). They are
\begin{eqnarray}
\label{eq:psix}
0&=&\nabla^2 \psi_y + \omega_y,\\
\label{eq:psiz}
0&=&\nabla^2 \psi_z + \omega_z,\\
\label{eq:omegax}
0&=&\nabla^2 \omega_y- i \beta\, \Ray\, \theta +\Ha_z^2 \left(-\partial_y\partial_z \phi  -\partial_z u_x+R\left(-\partial_y^2\phi+\omega_z-i\beta u_y\right)\right),\\
\label{eq:omegaz}
0&=&\nabla^2 \omega_z +\Ha_z^2 \left(-\partial_z^2 \phi +R\left(-\partial_y\partial_z\phi+\omega_y +i\beta u_z +R\partial_y u_x\right)\right).
\end{eqnarray}
The quantities $\omega_y$ and $\omega_z$ are the $y$- and $z$-components of the vorticity field $\nabla\times\bm{u}$. Equations for the remaining quantities are
 \begin{eqnarray}
\label{eq:theta}
0&=&\nabla^2 \theta +u_z,\\
\label{eq:phi}
0&=&\nabla^2 \phi -\omega_z -R \omega_y.
\end{eqnarray}
The last equation (\ref{eq:phi}) is obtained by substitution of Ohm's law (\ref{eq:Ohm}) into (\ref{eq:Ampere}). Combined with boundary conditions 
 on the top wall, bottom wall and side walls specified below, equations~(\ref{eq:psix})--(\ref{eq:phi}) represent a linear eigenvalue problem for the Rayleigh number $\Ray$ that must be solved numerically. A suitable discretization of this problem is obtained by a spectral collocation method with Chebyshev polynomials $T_n(z)=\cos\{n \arccos(z)\}$. The scalar fields such as $\theta$ are expanded as
\begin{equation}
\label{eq:doublechebyshev}
\theta(y,z)=\sum_{i}\sum_{k} \theta_{ik} \, T_i(2y/\Gamma)\, T_k(2z),
\end{equation}
where $-\Gamma/2\le y\le \Gamma/2$ and $-1/2 \le z\le 1/2$. The Poisson equations (\ref{eq:omegax}--\ref{eq:phi}) and boundary conditions are imposed pointwise at the Gauss-Lobatto collocation points
\begin{equation}
\label{eq:gausslobatto}
y_j=\Gamma\cos(j\pi/N_y)/2\quad (0\le j \le N_y),\qquad  z_k=\cos(k\pi/N_z)/2 \quad (0\le k \le N_z),
\end{equation}
where $N_y+1$ and $N_z+1$ are the number of expansion terms with respect to $y$ and $z$.

The boundary conditions for the vector stream function and vorticity components are determined with the help of equations (\ref{eq:velocitycomponents1})-(\ref{eq:velocitycomponents3}). Zero normal velocity on the horizontal walls requires $\psi_y=0$ and $\partial_z \psi_z=0$. On the $y$=$\pm\Gamma/2$ sidewalls, the corresponding conditions are $\psi_z=0$ and $\partial_y\psi_y=0$. The tangential velocity vanishes  on the sidewalls if $\omega_y=0$ and $\partial_y\psi_z=\partial_z\psi_y$.  On the top and bottom  walls these conditions are  $\omega_z=0$ and $\partial_y\psi_z=\partial_z\psi_y$.
The remaining boundary conditions for (\ref{eq:theta}) are $\theta=0$ on the top and bottom walls and $\partial_y\theta=0$ on the $y$=$\pm\Gamma/2$ sidewalls. The boundary condition for the electric potential  supplementing equation  (\ref{eq:phi}) is the homogeneous Neumann condition.

Since the boundary conditions (zero normal velocity) for (\ref{eq:psix}) and (\ref{eq:psiz})  only involve $\psi_y$ and $\psi_z$, respectively, one can represent the expansion coefficients of $\psi_y$ and $\psi_z$  by linear invertible maps through those of $\omega_y$ and $\omega_z$ (assuming the latter are augmented by the zero boundary values to be imposed on either $\psi_y$, $\psi_z$ or its normal derivatives). The expansions for $\psi_y$ and $\psi_z$ therefore contain $N_y+3$ and $N_z+3$ terms, respectively. The values of $\psi_y$, $\psi_z$ or its derivatives in equations~(\ref{eq:omegax})--(\ref{eq:phi})  (and associated boundary conditions) at the collocation points are represented through expansion coefficients of $\omega_y$ and $\omega_z$ via these linear invertible maps. The same can be done for the electric potential, which is the sum of a contribution from $\omega_z$ and $\omega_y$.   As a result of the collocation approximation one obtains a vector $\bm{Y}$ of unknowns containing the expansion coefficients of $\omega_y$, $\omega_z$, $\theta$  with a size of $3 (N_y+1)(N_z+1)$ and a generalized linear eigenvalue problem
\begin{equation}
\label{eq:generalizedevp}
\bm{\mathcal{A}} \bm{Y} =\Ray\, \bm{\mathcal{B}} \bm{Y}. 
\end{equation}
The method was implemented in Matlab \citep{MATLAB} using the default double precision. Notice that $\omega_y$, $\omega_z$ and $\phi$ are real variables. According to equations (\ref{eq:velocitycomponents1}-\ref{eq:velocitycomponents3}) and (\ref{eq:omegax}-\ref{eq:theta}), $u_y$, $u_z$ and $\theta$ would  be purely imaginary quantities. They are considered as real variables in the code and below. Problem (\ref{eq:generalizedevp}) was solved with Matlab's {\em eig} routine to find all eigenvalues and eigenvectors. The routine also works with a matrix $\bm{\mathcal{B}}$ whose rank is smaller than the rank of $\bm{\mathcal{A}}$ (as it is the case for (\ref{eq:generalizedevp})). It associates the spurious solutions that stem from equations not containing the eigenvalue $Ra$ with infinite eigenvalues. 

For the cases studied in this paper, the numerical resolution was typically $N_y=70$, $N_z=60$ for $Ha_z=50$ and $N_y=90$,  $N_z=70$
for $Ha_z=100$ with aspect ratio $\Gamma=2$.  A decrease  of $N_y$ by 10 typically only resulted in a relative change of the first  eigenvalue below $10^{-5}$. The generation of the matrices and the solution of problem (\ref{eq:generalizedevp}) for a given wavenumber took about 20 hours for $N_y=90$,  $N_z=70$ and about 6 hours for  $N_y=70$, $N_z=60$ on an Intel Xeon E5  CPU.

\subsection{Direct numerical simulations} \label{sec:DNS}
We conduct direct numerical simulations (DNS) of our magnetoconvection setup 
 using a second-order finite difference code developed by \citet{Krasnov:CF2011}. 
The governing equations are made dimensionless by using the cell height $H$, the imposed temperature difference $\Delta$, and the free-fall velocity $U = \sqrt{\alpha g \Delta H}$ (where $g$ and $\alpha$ are respectively the gravitational acceleration and the volumetric coefficient of thermal expansion of the fluid). The following non-dimensional equations are employed for our DNS:
\begin{eqnarray}
\nabla \cdot \boldsymbol{u}&=&0 \label{eq:continuity_DNS} \\
\frac{\partial \boldsymbol{u}}{\partial t}  + \boldsymbol{u}\cdot \nabla \boldsymbol{u} &=& -\nabla p + T\hat{z}+ \sqrt{\frac{\Pran}{Ra}} \nabla^2 \boldsymbol{u}+ \Ha_z^2\sqrt{\frac{\Pran}{Ra}}(\boldsymbol{j} \times {\boldsymbol{B}}),
\label{eq:Momentum_DNS} \\
\frac{\partial T}{\partial t} + \boldsymbol{u} \cdot \nabla T &=& \frac{1}{\sqrt{Ra \Pran}}\nabla^2 T, \label{eq:T_energy_DNS} \\
\boldsymbol{j} &=& -\nabla \phi + (\boldsymbol{u} \times {\boldsymbol{B}}),
\label{eq:Current_DNS}\\
\nabla^2 \phi &=& \nabla \cdot (\boldsymbol{u} \times {\boldsymbol{B}}), 
\label{eq:Potential_DNS}
\end{eqnarray}
 
 We numerically solve equations (\ref{eq:continuity_DNS}) to (\ref{eq:Potential_DNS}). 
The Prandtl number $\Pran$ is chosen to be 0.025, which is the same as that of mercury. 
We choose two Hartmann numbers based on vertical magnetic field: $\Ha_z=100$ and $\Ha_z=200$. For these Hartmann numbers, the corresponding critical Rayleigh numbers ($\Ray_c$) for the case without 
horizontal walls obtained using the linear stability analysis of \citet{Chandrasekhar:book:Instability}  is:
\begin{equation}
    \Ray_{c,\infty} = 
    \begin{cases}
        1.245 \times 10^5, & \Ha_z=100, \\
        4.48 \times 10^5, & \Ha_z=200.
    \end{cases}
    \label{eq:Rac_Chandrasekhar}
\end{equation}
Since we are interested in the wall-attached convection regime, we choose the Rayleigh number to be slightly below the critical Rayleigh numbers given by (\ref{eq:Rac_Chandrasekhar}). 
Thus, the chosen Rayleigh numbers are $\Ray=10^5$ for the case of $\Ha_z=100$ and $\Ray=4\times 10^5$ for the case of $\Ha_z=400$.
For each $\Ha_z$, we vary $R$ from 0 to 3.

We choose the domain-size to be $\Gamma \times \Gamma \times H = 4 \times 4 \times 1$ and employ a grid-resolution ranging from $1200 \times 1200 \times 300$ points to $1800 \times 1800 \times 400$.
The horizontal walls are at $z=\pm 1/2$ and the sidewalls are at $x=\pm \Gamma/2$ and $y=\pm \Gamma/2$. 
The mesh is non-uniform with stronger clustering of the grid points near the top and bottom boundaries. The elliptic equations for pressure, electric potential, and the temperature are solved based on applying cosine transforms in $x$- and $y$-directions and using a tridiagonal solver in the $z$-direction. The diffusive term in the temperature transport equation is treated implicitly. The time discretization of the momentum equation uses the fully explicit Adams-Bashforth/Backward-Differentiation method of second
order \citep{Peyret:book}. A constant time step size ranging from $5 \times 10^{-5}$ to $1 \times 10^{-4}$ free fall time unit was chosen for our simulations, which satisfied the Courant–Friedrichs–Lewy (CFL) condition for our runs. 

All the walls are rigid and electrically insulated such that the electric current density $\boldsymbol{j}$ forms closed field lines inside the cell. The top and bottom walls are held fixed at $T=-0.5$ and $T=0.5$ respectively, and the sidewalls are adiabatic with $\partial T/\partial \eta =0$ (where $\eta$ is the component normal to sidewall). All the simulations are initialized with the linear conduction profile for temperature (which is a function of the $z$-coordinate only) and a random noise of amplitude $A=0.001$ along the $z$-direction for velocity. We run the simulations initially on a coarse grid of $120 \times 120 \times 30$ points for 100 free-fall time units in which they converge to a statistically steady state. Following this, we successively refine the mesh to the required resolutions and  allow the simulations to converge after each refinement. Once the simulations reach the statistically steady state at the highest resolution, they are run for another 20 to 21 free-fall time units and a snapshot of the flow field is saved after every free-fall time unit. 

Since all the walls are no-slip, thin velocity boundary layers are formed adjacent to the walls.
For our simulations to be well-resolved, an adequate number of gridpoints need to be present in these boundary layers. 
It must be noted that the boundary layer profiles are strongly influenced by the magnetic fields.
For a purely vertical magnetic field, these boundary layers are categorized into Hartmann layers adjacent to the top and bottom walls and Shercliff layers adjacent to the sidewalls.
The thickness of Hartmann layers is given by $\delta_H=1/\Ha_z$ and that for the Shercliff layers is given by $\delta_S=1/\sqrt{\Ha_z}$.
However, in our case, the magnetic field is inclined with respect to the vertical direction; therefore, both the horizontal walls and $y$=$\pm\Gamma/2$ sidewalls will have a mix of Hartmann and Shercliff layers.  On the other hand, the $x$=$\pm\Gamma/2$ sidewalls will have purely Shercliff layers. 
Thus, we use $\delta$ only for representing the boundary-layer thickness.
For a conservative analysis, we estimate the thicknesses of these boundary layers as follows.
\begin{equation}
    \delta =
    \begin{cases}
         \min(\frac{1}{\Ha_z},\frac{1}{\sqrt{\R\Ha_z}}), & \mathrm{Horizontal~walls},    \\
         \min(\frac{1}{\sqrt{\Ha_z}},\frac{1}{\R\Ha_z}), & y\mbox{=}\pm\Gamma/2~\mathrm{sidewalls}, \\
         \dfrac{1}{(\Ha_z^2(1+R^2))^{1/4}}, & x\mbox{=}\pm\Gamma/2~\mathrm{sidewalls}.
    \end{cases} 
\end{equation}
Table~\ref{table:Simulation} lists the important parameters of our simulation runs. In this table, we also report the number of points in the different velocity boundary layers. We ensure that a  minimum of 10 points is present in the boundary layers so as to adequately resolve our simulation runs.

In the next section, we will discuss the results obtained from our stability analysis and numerical simulations.

\begin{table}
  \begin{center}
\def~{\hphantom{0}}
  \begin{tabular}{lccccccc}
      Runs  & $\Ha_z$   &  $\Ray$ & $R$ & Grid size & $n_{z}$ & $n_{y}$ & $n_{x}$ \\[3pt]
      1 & 100 & $1 \times 10^5$ & 0 & $1200 \times 1200 \times 300$ & 22 & 89 & 89 \\
      2 & 100 & $1 \times 10^5$ & 0.3 & $1200 \times 1200 \times 300$ & 22 & 39 & 88 \\
      3 & 100 & $1 \times 10^5$ & 1 & $1200 \times 1200 \times 300$ & 22 & 15 & 79 \\
      4 & 100 & $1 \times 10^5$ & 2 & $1200 \times 1200 \times 300$ & 22 & 15 & 120 \\
      5 & 100 & $1 \times 10^5$ & 3 & $1200 \times 1200 \times 300$ & 22 & 10 & 106 \\
      6 & 200 & $4 \times 10^5$ & 0 & $1200 \times 1200 \times 300$ & 12 & 21 & 21 \\
      7 & 200 & $4 \times 10^5$ & 0.3 & $1200 \times 1200 \times 300$ & 12 & 21 & 21 \\
      8 & 200 & $4 \times 10^5$ & 1 & $1200 \times 1200 \times 300$ & 12 & 13 & 18 \\
      9 & 200 & $4 \times 10^5$ & 2 & $1800 \times 1800 \times 400$ & 16 & 11 & 22 \\
      10 & 200 & $4 \times 10^5$ & 3 & $1800 \times 1800 \times 400$ & 16 & 12 & 18 \\
  \end{tabular}
  \caption{Parameters of the simulations: the Hartmann number ($\Ha_z$) based on the vertical magnetic field, the Rayleigh number ($\Ray$), the  ratio $R$ of the magnetic field strength along the horizontal to the vertical directions, the grid-size, the number of points in the velocity boundary layers along the horizontal walls ($n_{z}$), the $y$=$\pm\Gamma/2$ sidewalls ($n_{y}$), and the $x$=$\pm\Gamma/2$ sidewalls ($n_{x}$). The Prandtl number is constant in all cases at a value of $\Pran=0.025$.}
  \label{table:Simulation}
  \end{center}
\end{table}

\section{Results} \label{sec:Results}
In this section, we make a detailed analysis of the structure of the wall-modes and their impact on the heat and momentum transport. We will first present the linear stability analysis which is followed by a discussion of the results obtained from our direct numerical simulations. 
\subsection{Linear stability analysis} \label{sec:Stability_analysis}

In this subsection, we discuss the results from our linear stability model.
We assume that the set of finite positive eigenvalues of problem (\ref{eq:generalizedevp}) is sorted in ascending order, i.e. the smallest one will be referred to as the first eigenvalue etc.

\begin{figure}
  \centerline{\includegraphics[scale = 0.35]{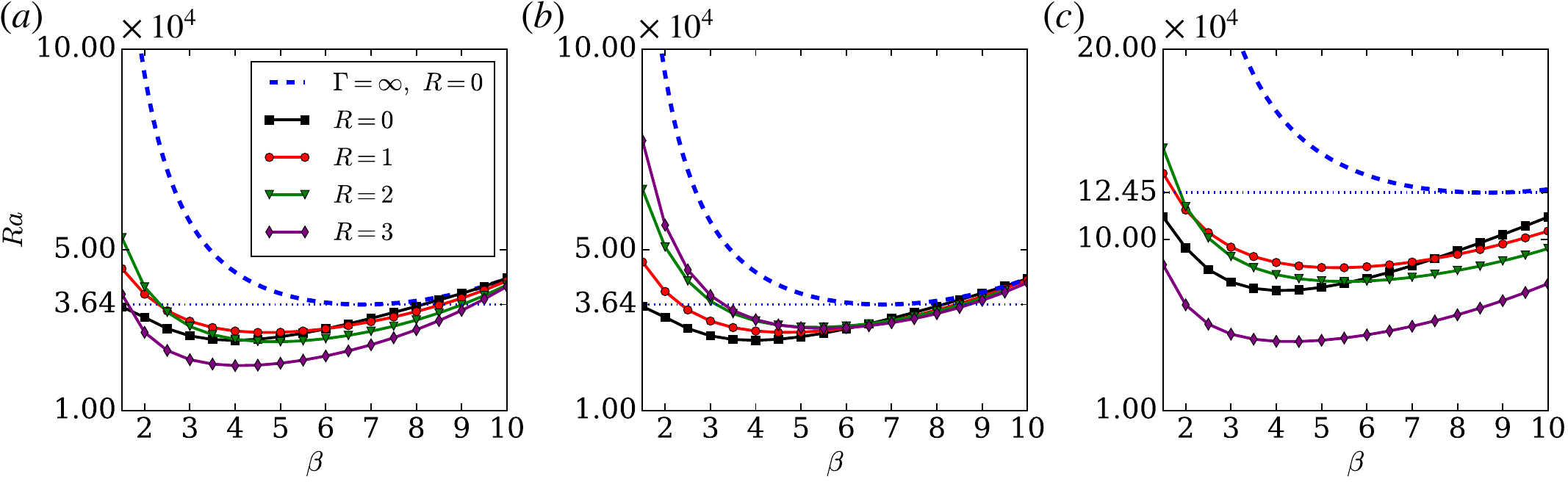}}
  \caption{Neutral stability curves for magnetoconvection with different values of $R$ for (a) $\Gamma=2$, $\Ha_z=50$; (b) $\Gamma=4$, $\Ha_z=50$; and (c) $\Gamma=2$, $\Ha_z=100$.}
\label{fig:Stability}
\end{figure}

\begin{figure}
  \centerline{\includegraphics[scale = 0.32]{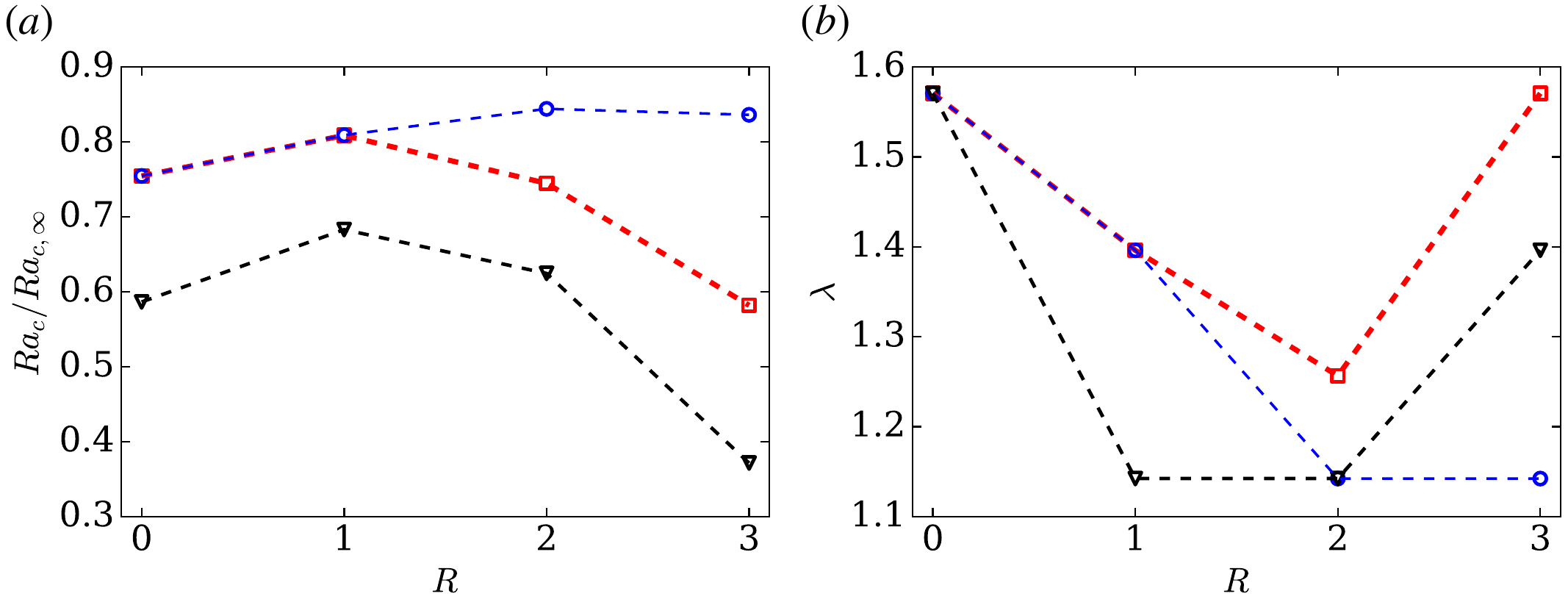}}
  \caption{For $\Ha_z=50$, $\Gamma=2$ (red squares); 
  $\Ha_z=50$, $\Gamma=4$ (blue circles); and
  $\Ha_z=100$, $\Gamma=2$ (black triangles): 
  (a) critical Rayleigh number $\Ray_c$, normalized with
  the same for infinite horizontal layer, and (b) the 
  wavelength of the most unstable mode at $\Ray_c$.
 }
\label{fig:Rac}
\end{figure}

In figures~\ref{fig:Stability}(a-c), we plot the first eigenvalue, denoted as Rayleigh number $\Ray$, at which the instability sets in due to disturbances at wavenumber $\beta$ for different values of $R$.
The minimum value of $\Ray$ is the critical Rayleigh number $\Ray_c$, and the wavenumber corresponding to $\Ray_c$ is the most unstable wavenumber.
 We also plot $\Ray$ for the infinite plane layer for the corresponding $\Ha_z$, where the flow is assumed to be uniform in the $y$-direction.
The figures show that $\Ray_c$ is smaller than that corresponding to the plane infinite layer ($\Ray_{c,\infty}$) for all $R$, $\Ha_z$, and aspect ratios considered in this study. 
This clearly implies that sidewalls destabilize the magnetoconvection system.
Figure~\ref{fig:Rac}(a) exhibits the dependence of $\Ray_c$ on $R$ for different aspect ratios and strengths of the vertical magnetic field.
It is evident from the above figure that for $R<1$, the critical Rayleigh number increases for all aspect ratios. 
For the larger aspect ratio box ($\Gamma=4$), $\Ray_c$ continues to increase with $R$ beyond $R=1$ and saturates at $R=2$. 
On the other hand, for the smaller aspect ratio box, the critical Rayleigh number starts to decrease with $R$ for $R>1$, a trend that does not seem to depend on $\Ha_z$. 

In figure~\ref{fig:Rac}(b), we exhibit the plots of the most unstable wavelength $\lambda$ versus $R$. 
The most unstable wavelength is calculated as 
$$\lambda= \frac{2\pi}{\beta_c},$$
where $\beta_c$ is the most unstable wavenumber.
The figure shows that $\lambda$, like $\Ray_c$, exhibits a non-monotonic variation with $R$ and lies between 1.1 and 1.6 for the entire range of parameters considered in our study. 
For small values of $R$, $\lambda$ decreases with $R$ and for the larger aspect-ratio box, it saturates at $R=2$. 
For the smaller aspect-ratio box, $\lambda$ increases with increasing $R$ beyond $R=2$. 

\begin{figure}
  \centerline{\includegraphics[scale = 0.35]{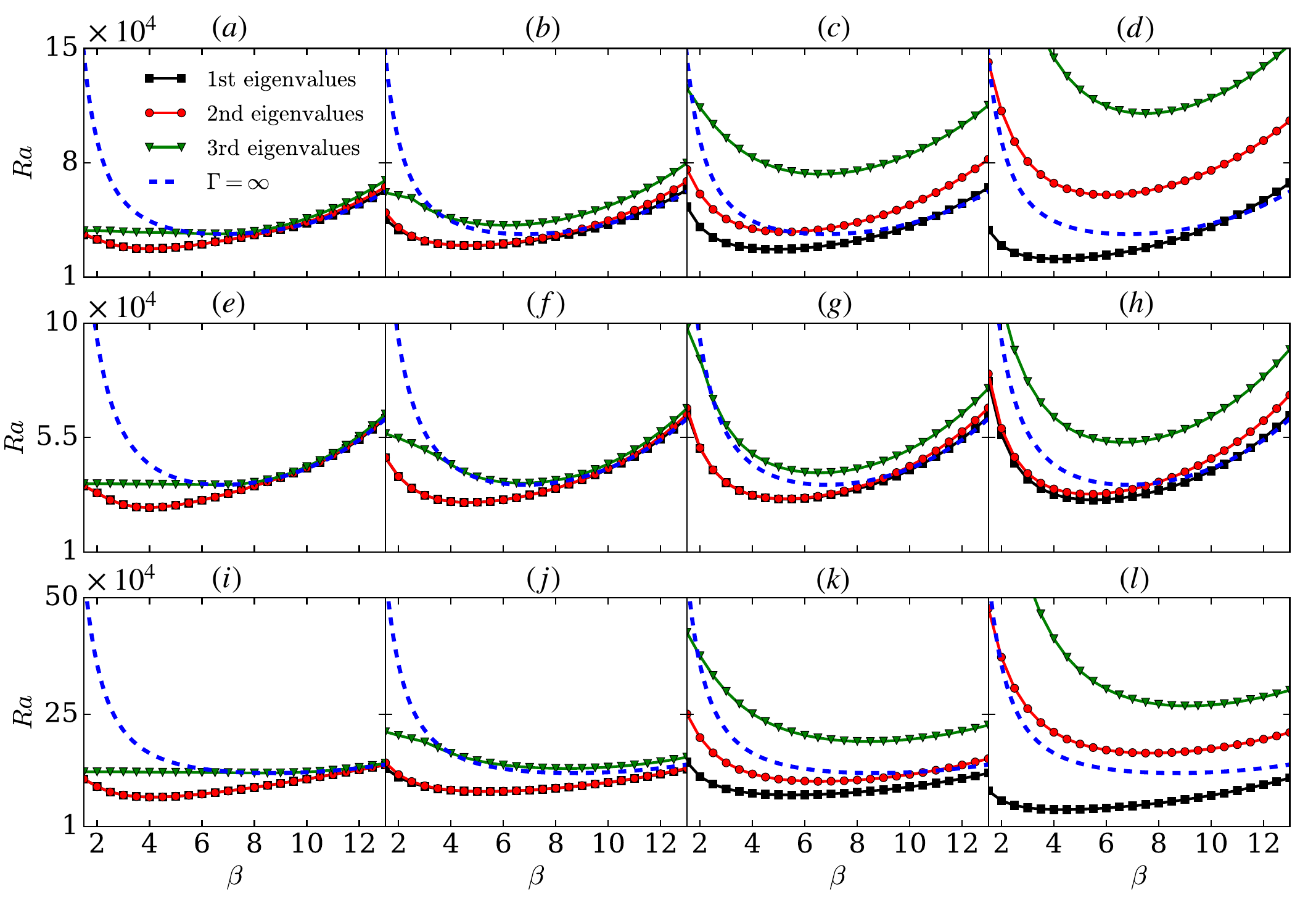}}
  \caption{
 Dependence of the first three eigenvalues on the wavenumber 
 $\beta$ for $\Gamma=2$ and $\Ha_z=50$ with (a) $R=0$, (b) $R=1$, (c) $R=2$, and (d) $R=3$; 
  $\Gamma=4$ and $\Ha_z=50$ with (e) $R=0$, (f) $R=1$, (g) $R=2$, and (h) $R=3$; and $\Gamma=2$ 
  and $\Ha_z=100$ with (i) $R=0$, (j) $R=1$, (k) $R=2$, and (l) $R=3$. Also shown is the neutral stability curve (dashed blue lines) for the corresponding infinite horizontal convection layer ($\Gamma=\infty$) with a purely vertical magnetic field ($R=0$).
  For the infinite plane layer, the flow is assumed to be uniform in the $y$-direction.
  }
\label{fig:Eigenmodes}
\end{figure}

We now examine the behavior of the first three eigenvalues ($\Ray$) for our cases.  
In figures~\ref{fig:Eigenmodes}(a)--(l), we plot the variations of these eigenvalues with $\beta$.
The neutral stability curve for the infinite plane convection layer for the corresponding $\Ha_z$ is also shown in dashed curves.
For $R=0$, it can be seen in  figures~\ref{fig:Eigenmodes}(a,e,i) that the minima of the third eigenvalue (represented by green triangles) overlaps with that of the plane convection layer. 
This indicates that the third eigenvalue corresponds to the onset of bulk convection for different wavenumbers. 
The figures also indicate that minima of the third eigenvalue, which corresponds to the critical Rayleigh number for convection in the bulk, increases with $R$.
This indicates that the bulk convection is further suppressed as the horizontal magnetic field increases.

\begin{figure}
  \centerline{\includegraphics[scale = 0.32]{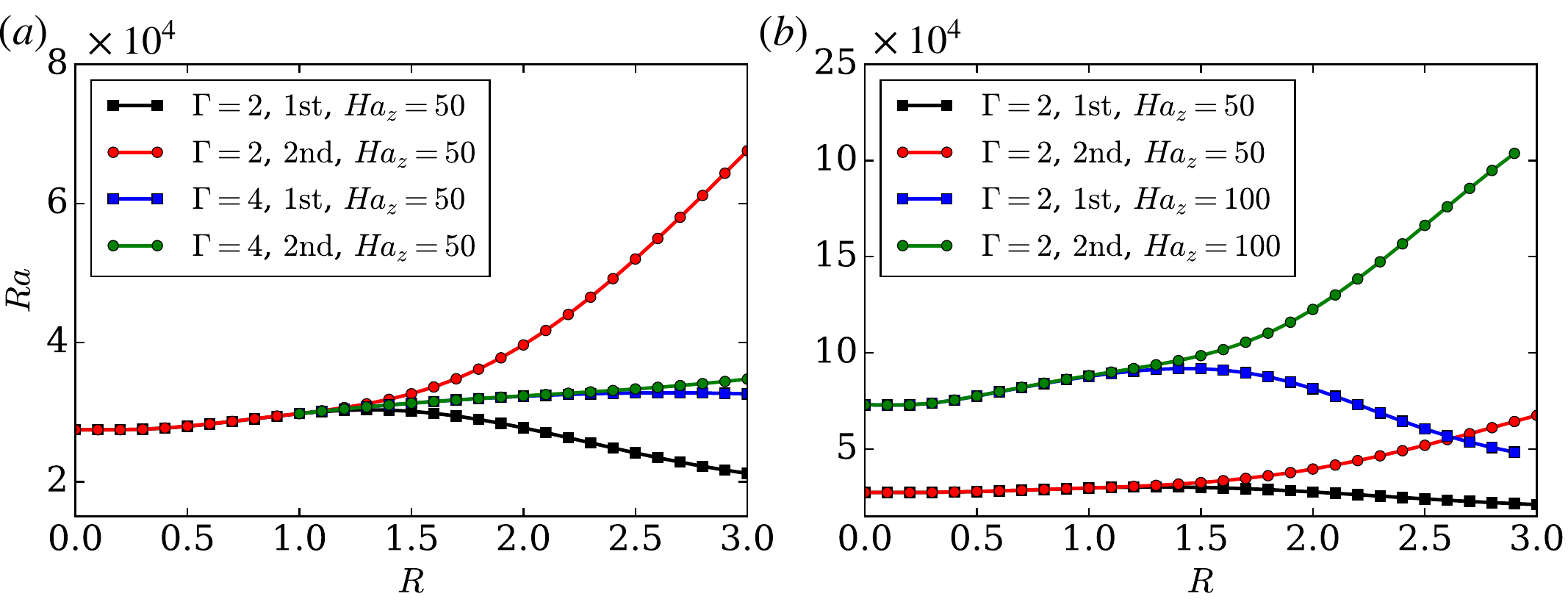}}
  \caption{Influence of (a) aspect ratio and (b) vertical Hartmann number on the stability curves for 1st and 2nd eigenvalues as function of the ratio of the horizontal to vertical magnetic fields at fixed wavenumber $\beta=4$.
 }
\label{fig:bxdependence}
\end{figure}

The first and second eigenvalues 
(denoted by black squares and red circles respectively) correspond to wall-attached convection, and their minima are less than that for the third eigenvalue (corresponding to bulk convection).  
These eigenvalues nearly overlap for small horizontal magnetic fields but begin to diverge at large horizontal magnetic fields.
These variations are shown more explicitly in figures~\ref{fig:bxdependence}(a,b) in which we exhibit the variations of the first two eigenvalues with $R$ for $\beta=4$ (near the most unstable wavenumber). 
It is evident from these figures that the eigenvalues diverge visibly for $R>R_c\approx 1.5$ for $\Gamma=2$ and for $R>R_c \approx 2.5$ for $\Gamma=4$. 
It is interesting to note that the value of $R$ for which the eigenvalues  begin to diverge does not seem to depend on $\Ha_z$. 
\begin{figure}
  \centerline{\includegraphics[scale = 0.35]{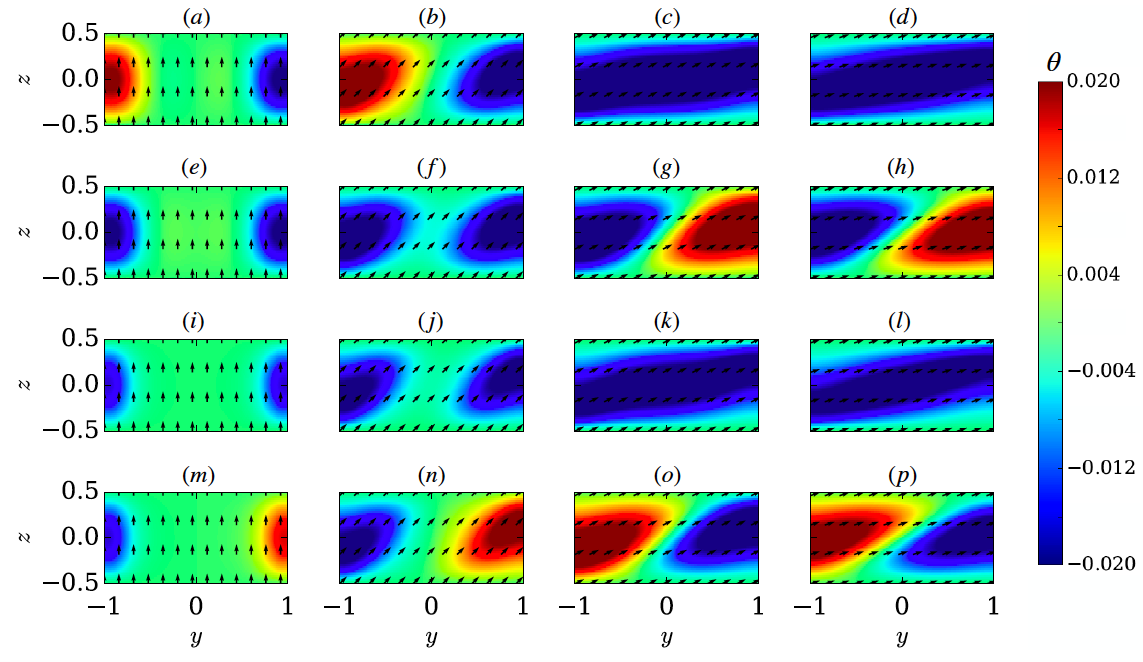}}
  \caption{Eigenvector field from the linear stability analysis. Contours of the temperature deviation $\theta$ are shown. For $\beta=4$, $\Gamma=2$, and $\Ha_z=50$: isocontours of $\theta$ corresponding to the first solution for (a) $R=0$, (b) $R=1$, (c) $R=2$, and (d) $R=3$. For $\beta=4$, $\Gamma=2$, and $\Ha_z=50$: isocontours of $\theta$ corresponding to the second solution for (e) $R=0$, (f) $R=1$, (g) $R=2$, and (h) $R=3$. For $\beta=4$, $\Gamma=2$, and $\Ha_z=100$: isocontours of $\theta$ corresponding to the first solution for (i) $R=0$, (j) $R=1$, (k) $R=2$, and (l) $R=3$. For $\beta=4$, $\Gamma=2$, and $\Ha_z=100$: isocontours of $\theta$ corresponding to the second solution for (m) $R=0$, (n) $R=1$, (o) $R=2$, and (p) $R=3$.  Also shown are the vector plots of the corresponding magnetic fields.
 }
\label{fig:eigenmodes_contours}
\end{figure}
\begin{figure}
  \centerline{\includegraphics[scale = 0.18]{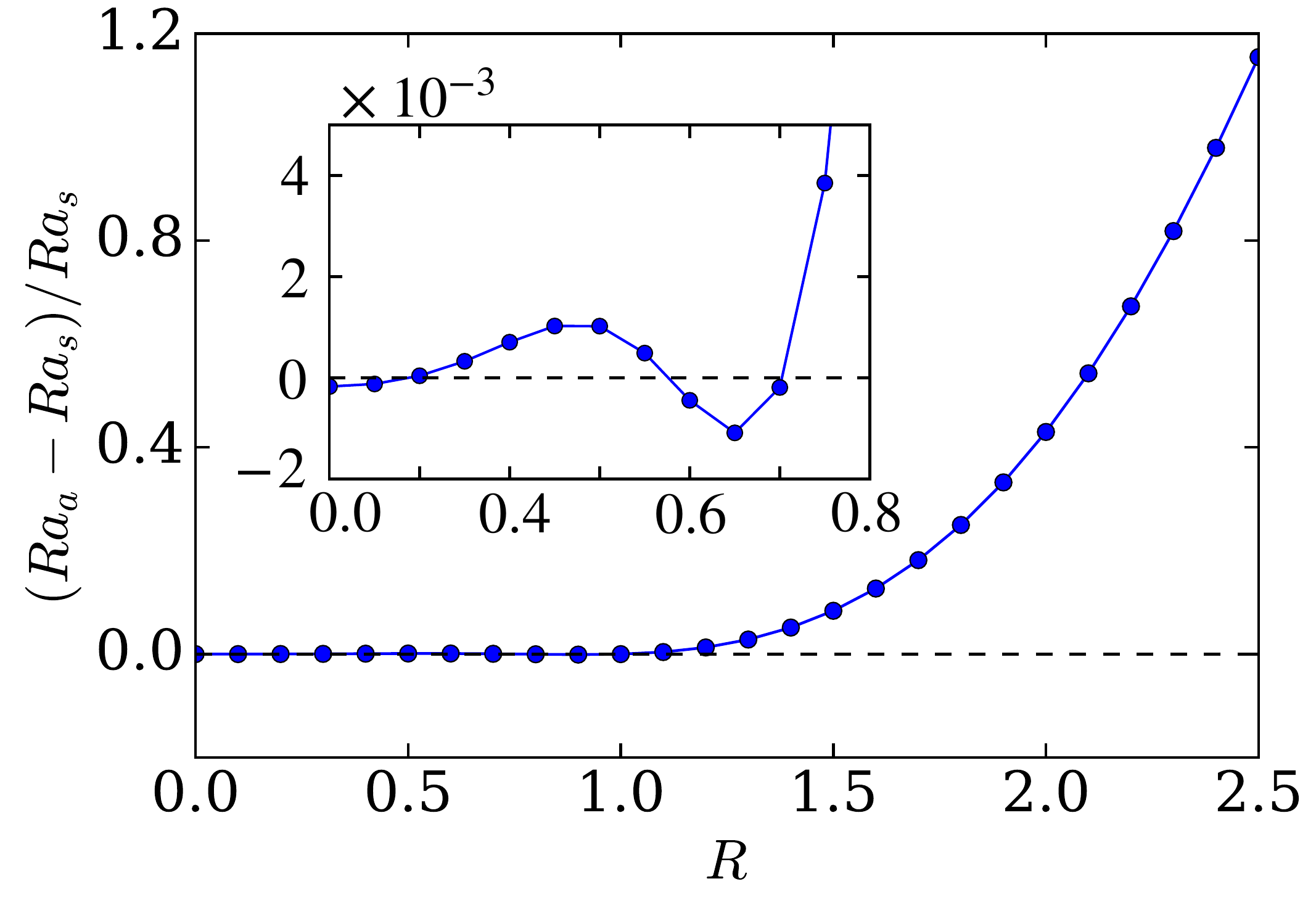}}
  \caption{Ratio of the difference between the critical Rayleigh numbers for antisymmetric solution ($\Ray_a$) and the symmetric solution ($\Ray_s$) to the critical Rayleigh number of the symmetric solution. The inset exhibits a magnified version of the above plot with $R$ ranging from 0 to 0.8. Parameters are $Ha_z=50$, $\Gamma=2$ and $\beta=4$.  
 }
\label{fig:Sym_asym}
\end{figure}
\begin{figure}
  \centerline{\includegraphics[scale = 0.34]{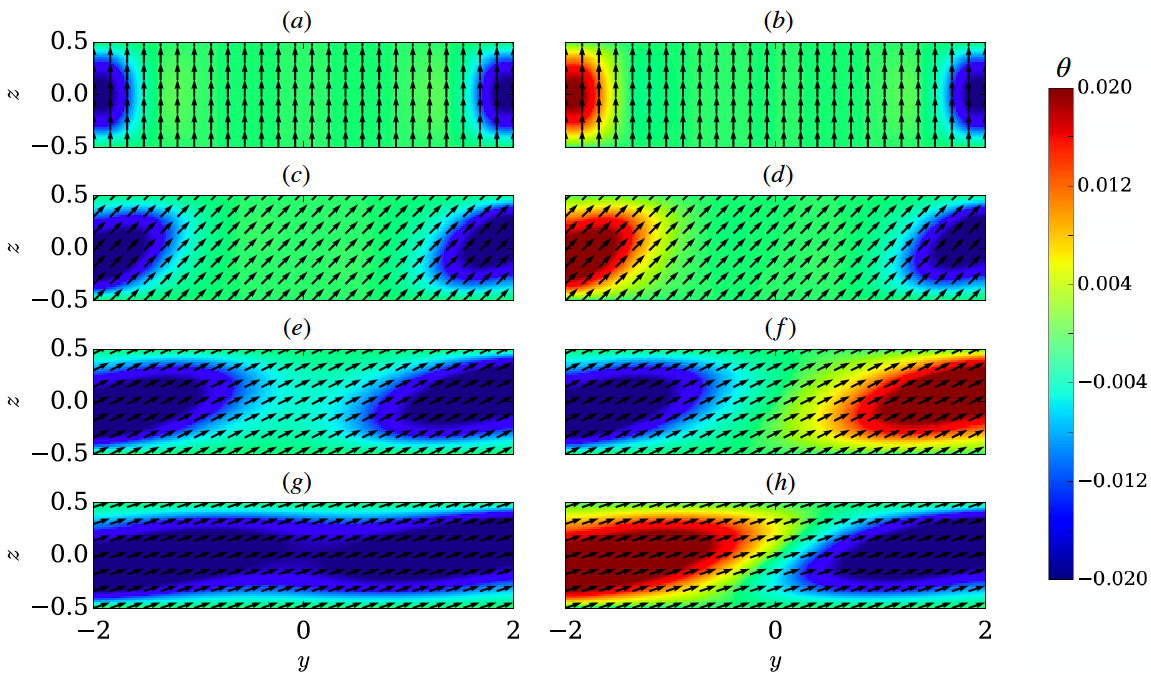}}
  \caption{Eigenvector field from the linear stability analysis. Contours of the temperature deviation $\theta$ are shown. For $\Ha_z=50$, $\beta=4$, and $\Gamma=4$: isocontours of $\theta$ corresponding to the first eigensolutions for (a) $R=0$, (c) $R=1$, (e) $R=2$, and (g) $R=3$, and corresponding to second eigensolutions for (b) $R=0$, (d) $R=1$, (f) $R=2$, and (h) $R=3$. Also shown are the vector plots of the corresponding magnetic fields.}
\label{fig:eigenmodes_contours_L4}
\end{figure}

We will now examine the spatial structure of the  wall-modes, i.e. the  eigenfunctions that
correspond to the first and second eigenvalues.
The eigenfunctions are normalized such that the maximum of the vertical velocity becomes equal to unity.

In figures~\ref{fig:eigenmodes_contours}(a)--(p), we exhibit the contour plots of the temperature perturbation $\theta$ on the vertical $y$-$z$ plane corresponding to the first and second eigenvalues for $\beta=4$, $\Gamma=2$, and different $\Ha_z$ and $R$.
The first and third rows of the figure correspond to the first eigensolutions for $\Ha_z=50$ and $\Ha_z=100$, respectively, whereas the second and fourth rows correspond to the second eigensolutions.
The figures show that as the horizontal magnetic field strength is increased, the wall modes get elongated and tilt along the direction of the resultant magnetic field. 
In fact, for $R\geq 2$, the wall modes are no longer confined adjacent to the sidewalls; instead, they occupy almost the entire bulk and interact with each other as explained later in this section.

Figures~\ref{fig:eigenmodes_contours}(a)--(p) also show that two there are two types of spatial structures corresponding to the first two eigensolutions. 
The first type consists of a hot plume ($\theta>0$) adjacent to the $y=-\Gamma/2$ sidewall and a cold plume ($\theta<0$) adjacent to $y=\Gamma/2$ sidewall; the corresponding eigensolution will be referred to as {\em antisymmetric solution}, see panel (a).
The second type consists of cold (or hot) plumes adjacent to both the sidewalls; the corresponding eigensolution will be referred to as {\em symmetric solution}, see panel (e). 
For $R<1.5$, there is only a marginal difference between  the symmetric ($\Ray_s$) and antisymmetric ($\Ray_a$) eigenvalues. As exhibited in figure~\ref{fig:Sym_asym}, this difference is less than 0.5\% for $R<0.8$ and is approximately $10\%$ at $R=1.5$ for the case of $\Ha_z=50$ and $\Gamma=2$.
Thus, for $R<1.5$, there is nearly an equal preference for symmetric and antisymmetric structures to develop at the onset of convection. 
However, these eigenvalues deviate significantly once  R exceeds the threshold $R_c \approx 1.5$,  above which the eigenvalue corresponding to the symmetric solution becomes significantly smaller. 
This implies that there is a stronger preference for the symmetric structures to develop at  the onset of convection for $R>R_c$.
In this regime of $R$, the symmetric eigensolution comprises of a large plume developed by the merging of two cold (or hot) plumes adjacent to the opposite walls as visible in figures~\ref{fig:eigenmodes_contours}(c), (d), (k), and (l).
On the other hand, as seen in figures~\ref{fig:eigenmodes_contours}(g), (h), (o), and (p), the antisymmetric eigensolutions for $R > R_c$ comprise of a cold plume extending on the top of a hot plume (or vice-versa) extended from the opposite wall, resulting in two convection rolls on top of each other.     
Our observations imply that a clear separation of the symmetric and antisymmetric solutions occurs when the wall modes adjacent to the opposite walls for symmetric solutions start interacting with each other.
The merging of the plumes and the resultant formation of a merged roll results in an increased heat and momentum transport and hence in the decrease of the critical Rayleigh number. 

Figures~\ref{fig:eigenmodes_contours_L4}(a--h) exhibit the contour plots of $\theta$ on vertical $y$-$z$ midplane corresponding to the first and second eigensolutions for $\Gamma=4$, $\beta=4$, and $\Ha_z=50$. 
These figures again show that the wall modes tend to elongate along the direction of the resultant magnetic field. 
Similar to the $\Gamma=2$ cases, the two eigenvalues correspond to symmetric and antisymmetric solutions respectively.
The wall modes are symmetric for the first eigensolution and antisymmetric for the second. 
It can be recalled from figure~\ref{fig:bxdependence}(a) that the first and second eigenvalues for $\Gamma=4$ box begin to diverge at a higher value of $R$ compared to the $\Gamma=2$ box. 
A clear separation occurs near $R_c \approx 2.5$; this is because owing to the larger aspect ratio of the box, the plumes adjacent to opposite walls are able to interact and merge only at higher tilts, and hence at larger $R$. 
The merged plume is exhibited in figure~\ref{fig:eigenmodes_contours_L4}(g) which displays the contours of $\theta$ corresponding to the symmetric solution for $R=3$. 

Our analysis suggests that the nonmonotonic behaviour of $\Ray_c$ and wavelength $\lambda$ with respect to $R$ is due to the increasing interaction 
of the plumes on opposite sidewalls. 
As long as the opposite plumes do not merge, an increase in the horizontal magnetic field component further stabilizes the magnetoconvection system with $\Ray_c$ increasing and the wavelength $\lambda$ decreasing with $R$. 
The system gets destabilized due to the merging of the opposite plumes and the variations of $\Ray_c$ and $\lambda$ with $R$ get reversed. In the next subsection, we will discuss the results of direct numerical simulations of the nonlinear evolution of magnetoconvection.

\subsection{Results of direct numerical simulations} \label{sec:DNS_Results}
In this subsection, we analyze our steady-state DNS results and examine the structures of the wall modes, their role in the heat and momentum transport, and the effects of initial conditions on the formation of the wall modes.

\subsubsection{Structure of the wall modes} \label{sec:Wall_modes_structure}
We use our numerical data to study the spatial convection structures for $\Ha_z=100$ and $\Ha_z=200$ with $R$ ranging from 0 to 3.

\begin{figure}
  \centerline{\includegraphics[scale = 0.2]{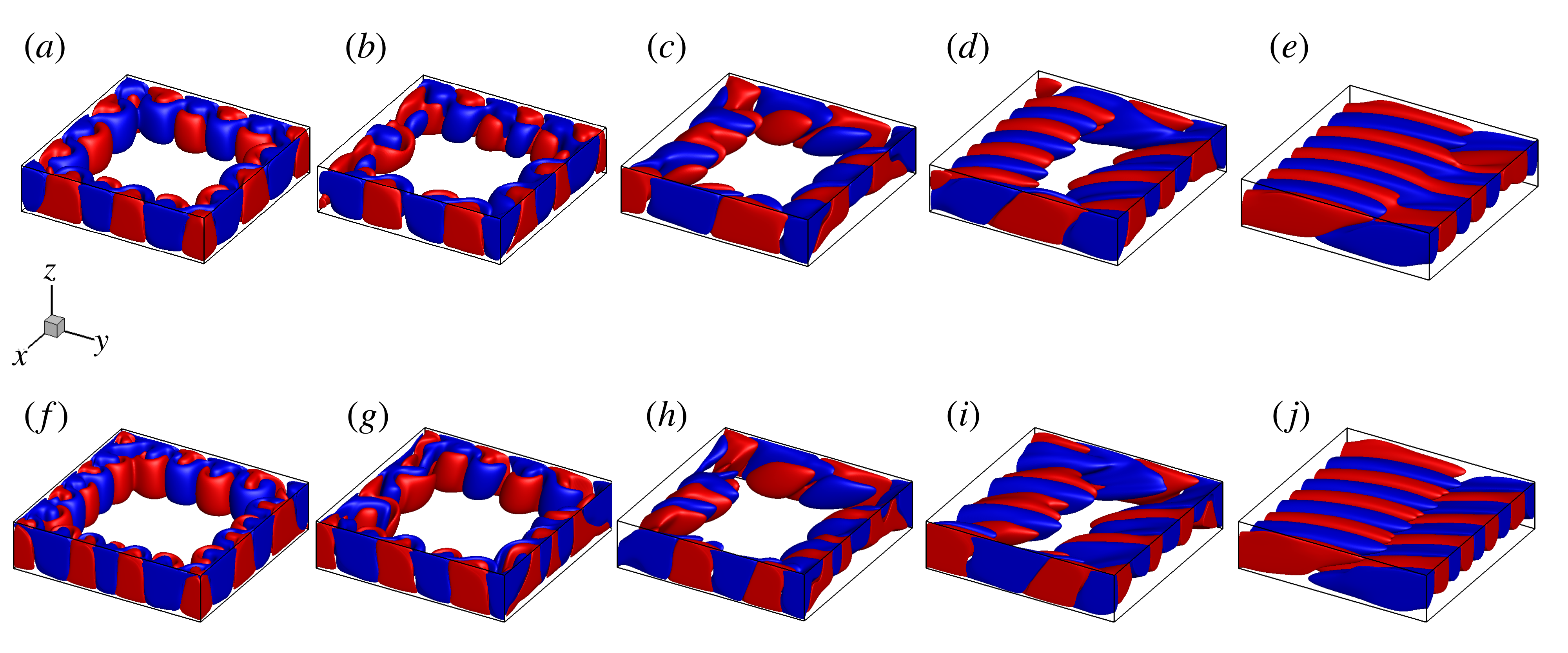}}
  \caption{Results of direct numerical simulations for the vertical velocity component. Isosurfaces of $u_z=0.01$ (red) and $u_z=-0.01$ (blue) for $Ha_z=100$ with (a) $R=0$, (b) $R=0.3$, (c) $R=1$, (d) $R=2$, and (e) $R=3$, and for $\Ha_z=200$ with (f) $R=0$, (g) $R=0.3$, (h) $R=1$, (i) $R=2$, and (j) $R=3$.}
\label{fig:Wall_modes_isosurface}
\end{figure}

\begin{figure}
  \centerline{\includegraphics[scale = 0.4]{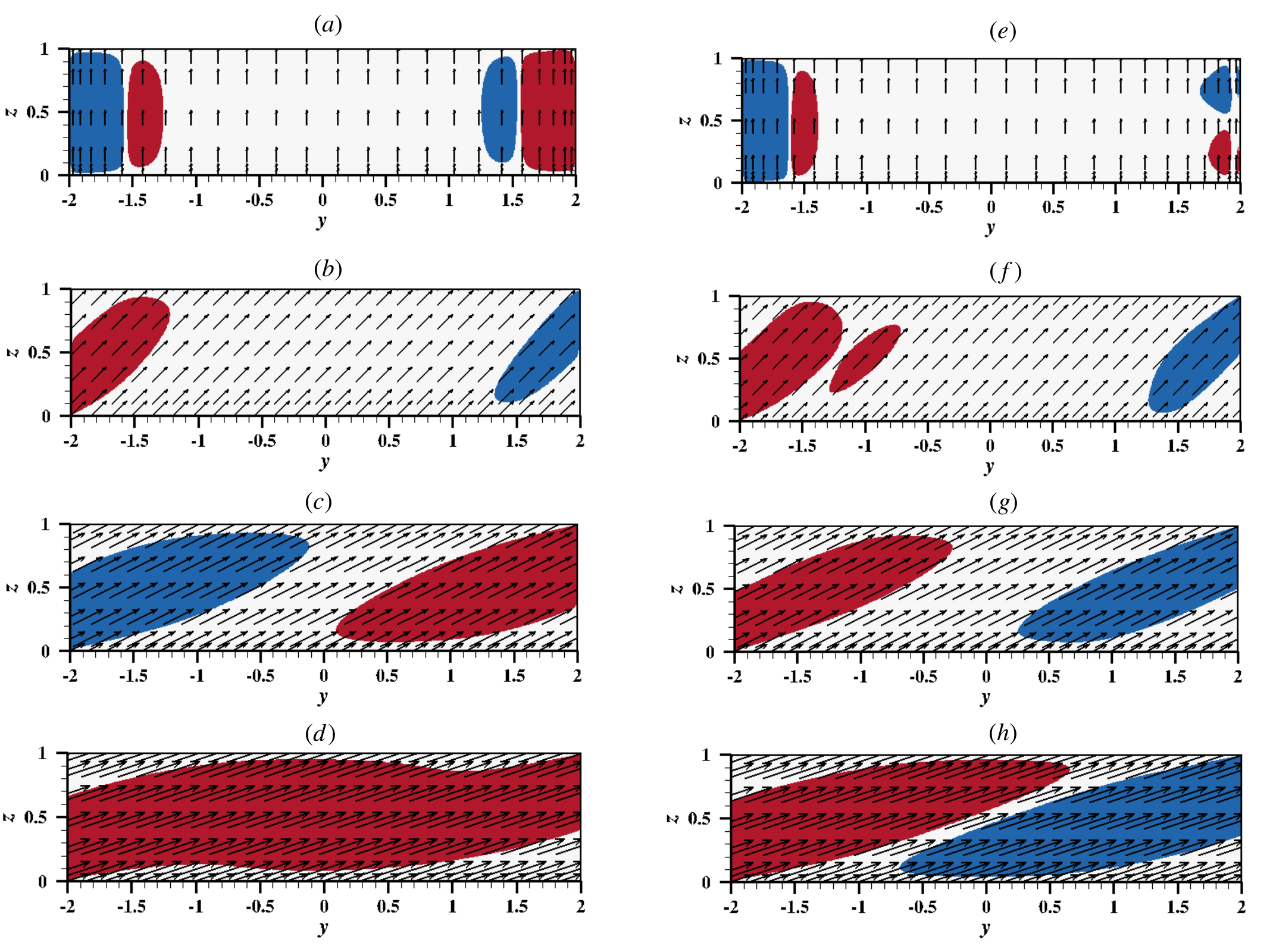}}
  \caption{Results of direct numerical simulations for the wall modes. Contours of $u_z \geq 0.01$ (red) and $u_z \leq -0.01$ (blue) along with the vector plots of $\boldsymbol{B}$ on $x=0$ midplane for $Ha_z=100$ with (a) $R=0$, (b) $R=1$, (c) $R=2$, (d) $R=3$, and for $\Ha_z=200$ with (e) $R=0$, (f) $R=1$, (g) $R=2$, (h) $R=3$. The wall modes align themselves along the direction of $\boldsymbol{B}$.}
\label{fig:Contours_midplane}
\end{figure}

In figures~\ref{fig:Wall_modes_isosurface}(a)--(j), we exhibit the isosurfaces of $u_z = \pm 0.01 $ for all our runs.
In figures~\ref{fig:Contours_midplane}(a)--(h), we display the contours of $u_z \geq 0.01$ (red) and $u_z \leq 0.01$ (blue); these contours give a visualization of the upwelling and downwelling plumes respectively.
These figures show the presence of wall-attached convection with suppressed fluid flow in the bulk. 
Consistent with the results of the stability analysis in the previous subsection, these wall modes become elongated and align themselves along the direction of the resultant magnetic field as the horizontal magnetic field is increased.
Figures~\ref{fig:Wall_modes_isosurface} and \ref{fig:Contours_midplane} also show that with the exception of the case with $\Ha_z=100$ and $R=3$, the wall modes are antisymmetric, that is, the plumes are upwelling (or downwelling) adjacent to $y=-\Gamma/2$ wall and downwelling (or upwelling) adjacent to $y=\Gamma/2$ wall. 
The wall modes occupy the entire bulk for $R=3$, consistent with the stability analysis of the $\Gamma=4$ box. 
For the above field ratio, the wall modes for the $\Ha_z=100$ case are symmetric and the plumes adjacent to the opposite walls merge to form a large plume as displayed in figures~\ref{fig:Wall_modes_isosurface}(e) and \ref{fig:Contours_midplane}(d).

It can also be observed in figures~\ref{fig:Wall_modes_isosurface}(a--j) that there are slight irregularities in the wall-mode structures. These irregularities seem to be associated with small temporal fluctuations ($\sim 1\%$) in integral quantities that remain after our simulations reach an apparently stationary state. 
These fluctuations indicate that the wall modes are still evolving, albeit slowly.
However, it is important to note that we could not observe any noticeable change in the structures of the wall modes over the time-span of our simulations. Any change in the spatial arrangement of the modes is likely to be visible only after the simulations are run for several times the diffusion time scale, which is around 150 to 200 free-fall time units.
Hence our solutions can be considered to be effectively steady.

We estimate the wavelength of the wall modes by visual inspection of the vertical velocity isosurfaces in figures~\ref{fig:Wall_modes_isosurface}(a)--(j). 
We consider only those wall modes that are adjacent to $y=\pm \Gamma/2$ sidewalls; these walls are not parallel to the resultant magnetic field.
The estimation of the wavelength is done as follows. 
We count the number of upwelling (or downwelling) plumes adjacent to both $y=-\Gamma/2$ and $y=\Gamma/2$ sidewalls and calculate their average as $N_{p}$. The wavelength $\lambda_{wm}$ of the wall modes is given by 
\begin{equation}
    \lambda_{wm} = \frac{\Gamma}{N_p}.
    \label{eq:Wavelength_estimation}
\end{equation}
\begin{figure}
  \centerline{\includegraphics[scale = 0.4]{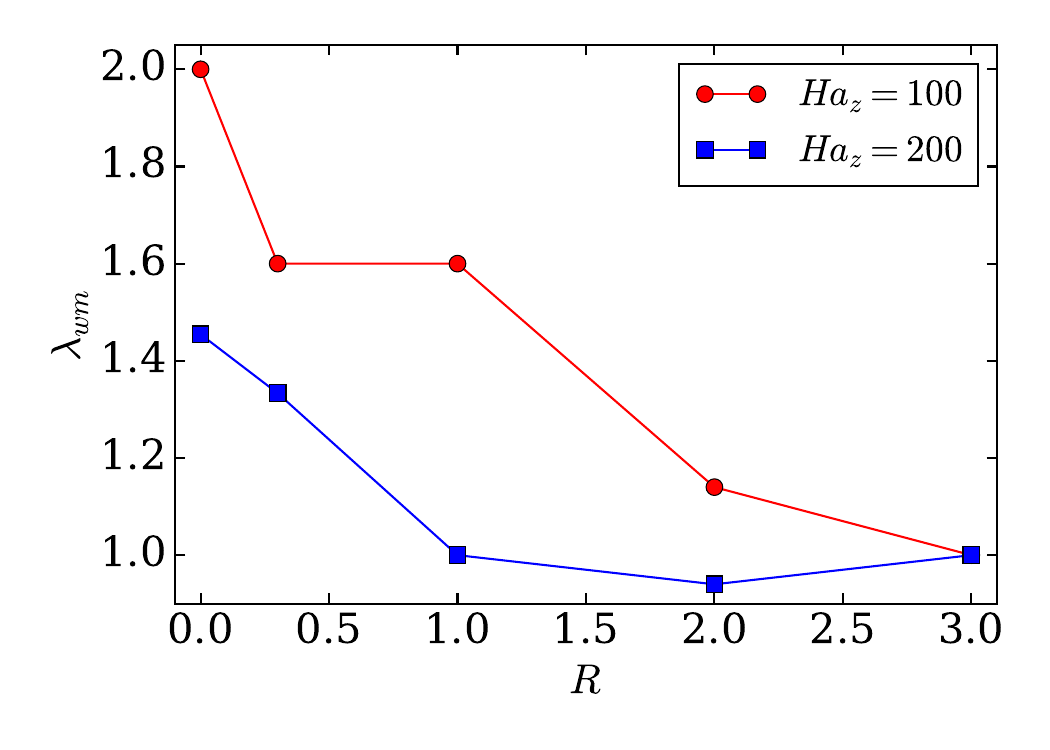}}
  \caption{Variations of the wavelength $\lambda_{wm}$ of the wall modes with $R$. The wavelength tends to decrease with an increase in the horizontal magnetic field. 
 }
\label{fig:Wavelength_DNS}
\end{figure}
We plot the estimated $\lambda_{wm}$ versus $R$ in figure~\ref{fig:Wavelength_DNS}.
The figure shows that $\lambda_{wm}$ tends to decrease as the horizontal magnetic field increases and saturates at large $R$.
Further, the wavelength corresponding to $Ha_z=200$ is smaller than that for $\Ha_z=100$ except for $R=3$. 
This is in line with the fact that the threshold wavelength decreases with increasing vertical magnetic field strength~\citep{Chandrasekhar:book:Instability}. 
It can be recalled that a similar trend was observed in the variation of the most unstable wavelength with $R$ in our stability analysis in \S~\ref{sec:Stability_analysis}.
However, it must be kept in mind that since the Rayleigh numbers considered in our DNS are higher than the threshold Rayleigh number above which the wall modes appear, there is always a chance for secondary modes with smaller wavenumbers to develop.

\subsubsection{Heat and momentum transport} \label{sec:Heat_Transport}
\begin{figure}
  \centerline{\includegraphics[scale = 0.35]{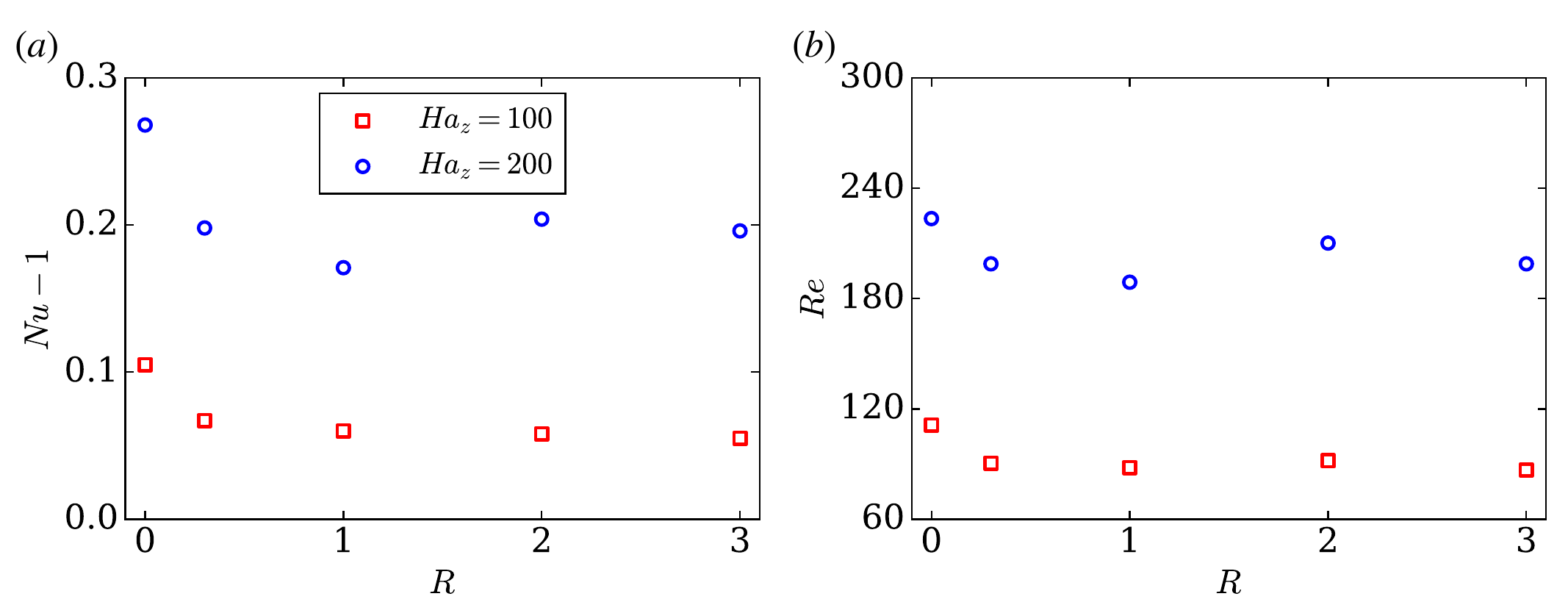}}
  \caption{Plots of (a) the nondimensional convective heat flux $\Nu-1$ and (b) the Reynolds number $\Rey$ versus $R$. For $R \leq 1$, the heat and momentum transport decreases as $R$ is increased.}
\label{fig:NuRe}
\end{figure}

We will now explore the influence of the wall modes on the heat and momentum transport.
We compute the Nusselt ($\Nu$) and the Reynolds numbers ($\Rey$) using our numerical data as follows: 
\begin{eqnarray}
    \Nu &=& 1 + \sqrt{\Ray \Pran}\langle u_zT \rangle_V, \label{eq:Nu} \\
    \Rey &=& \sqrt{\frac{\Ray}{\Pran}}U_{rms}, \label{eq:Re}
\end{eqnarray} 
where $U_{rms} = \sqrt{\langle u_x^2 + u_y^2 + u_z^2 \rangle_V}$ is the root mean square velocity and $\langle \cdot \rangle_V$ denotes volume averaging. 
The second term of the right hand side of (\ref{eq:Nu}) is the normalized convective heat flux.
We plot the normalized heat flux and the Reynolds number versus $R$ for all our runs in figures~\ref{fig:NuRe}(a) and (b) respectively.
The figures show that for $R<1$, both $\Nu$ and $\Rey$ decrease with $R$. 
This is consistent with our conclusion from the stability analysis in \S~\ref{sec:Stability_analysis} that for small values of $R$, an increase in the horizontal magnetic field results in the stabilization of the magnetoconvection system. 
There is, however, no clear trend in the variations of $\Nu$ and $\Rey$ for $R>1$.

\begin{figure}
  \centerline{\includegraphics[scale = 0.36]{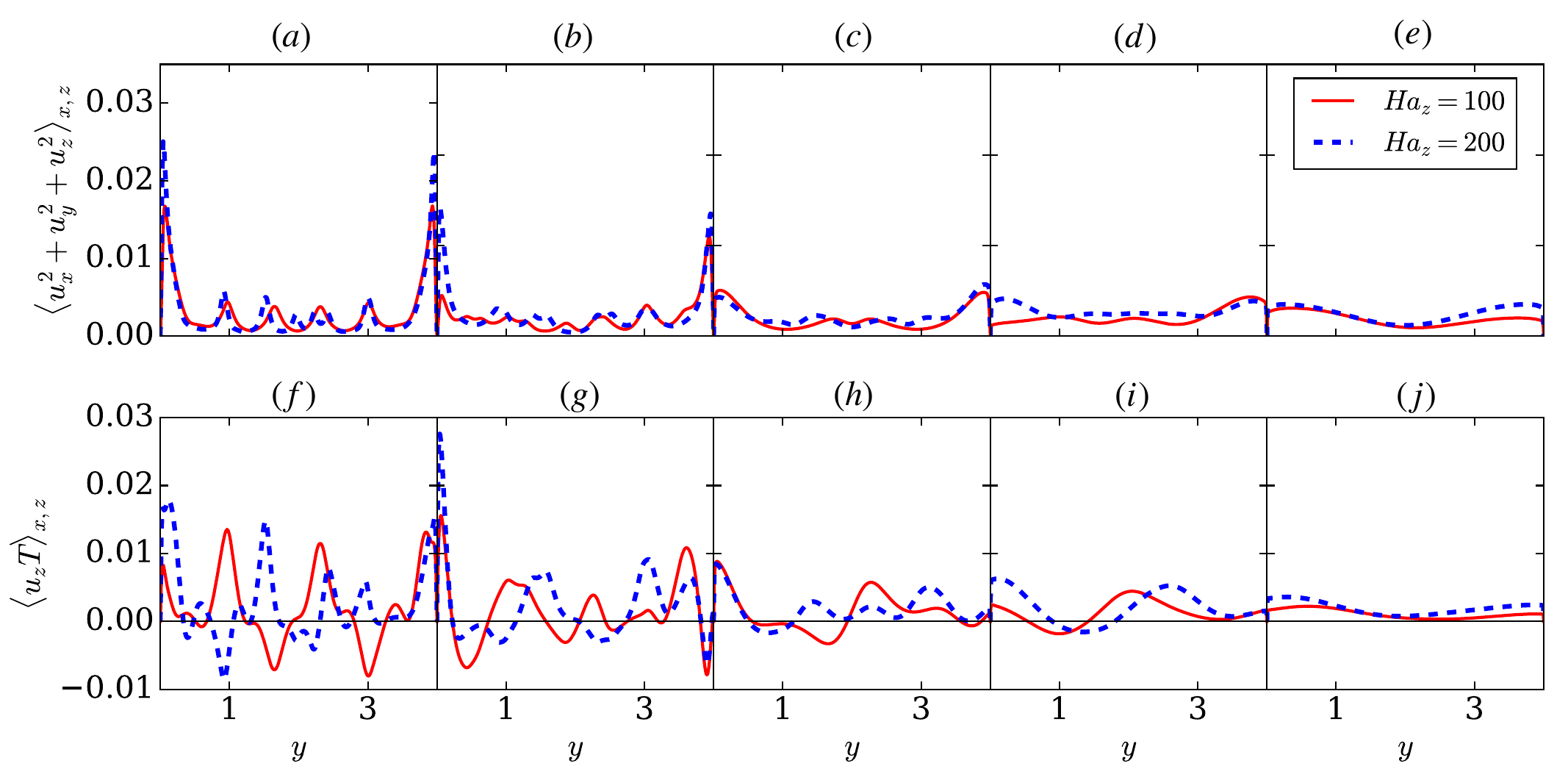}}
  \caption{Spatial distribution of kinetic energy $E$ and the convective heat flux $u_zT$. Plots of 2$\langle E \rangle_{x,z}=\langle u_x^2 + u_y^2 + u_z^2 \rangle_{x,z}$ for (a) $R=0$, (b) $R=0.3$, (c) $R=1$, (d) $R=2$, and (e) $R=3$. Plots of $\langle u_zT \rangle_{x,z}$ for (f) $R=0$, (g) $R=0.3$, (h) $R=1$, (i) $R=2$, and (j) $R=3$. In the above, $\langle \cdot \rangle_{x,z}$ represents averaging over $x$-$z$ plane.}
\label{fig:y_distribution}
\end{figure}

Having studied the trends of the global heat and momentum transport with $R$, we will now explore the spatial variation of heat and momentum transport for different regimes of $R$.
In figures~\ref{fig:y_distribution}(a)--(e), we plot twice the kinetic energy $E=0.5(u_x^2 + u_y^2 + u_z^2)$, averaged over the $x$-$z$ plane, versus $y$, which is the direction parallel to the horizontal magnetic field.
For $R=0$, there are two sharp peaks close to $y=-2$ and $y=2$ for both $\Ha_z=100$ and $\Ha_z=200$; these peaks correspond to the wall modes.
The bulk region lies in between these two peaks. The bulk consists of several smaller peaks  and the kinetic energy in this region is much less compared to the near-wall regions.
As $R$ is increased, the height of the near-wall peaks decreases and of those in the bulk increases. 
Thus, the distribution of kinetic energy along $y$ becomes more uniform as $R$ is increased. 
It is a known fact that in magnetohydrodynamic flows, the Lorentz force tends to modify the flow-field so as to minimize the gradients of velocity along the direction of the magnetic field~\citep{Davidson:book:MHD}.
Thus, in our case, as $R$ is increased, the Lorentz force generated due to the $y$-component of the magnetic field becomes strong and suppresses the gradients of velocity along $y$.

In figures~\ref{fig:y_distribution}(f)--(j), we plot the local convective heat flux $u_zT$, averaged over the $x$-$z$ plane, along $y$.
Although the volume-averaged heat flux is positive in thermal convection, the local heat flux for $R=0$ fluctuates between positive and negative values as one proceeds along $y$. 
These fluctuations even out as $R$ is increased, and for $R=3$, the local heat flux remains positive throughout with very small gradients along $y$.
Again, this is due to the strong Lorentz forces generated by the horizontal component of the magnetic field which suppresses the gradients of velocity along the magnetic field's direction.

\begin{figure}
  \centerline{\includegraphics[scale = 0.36]{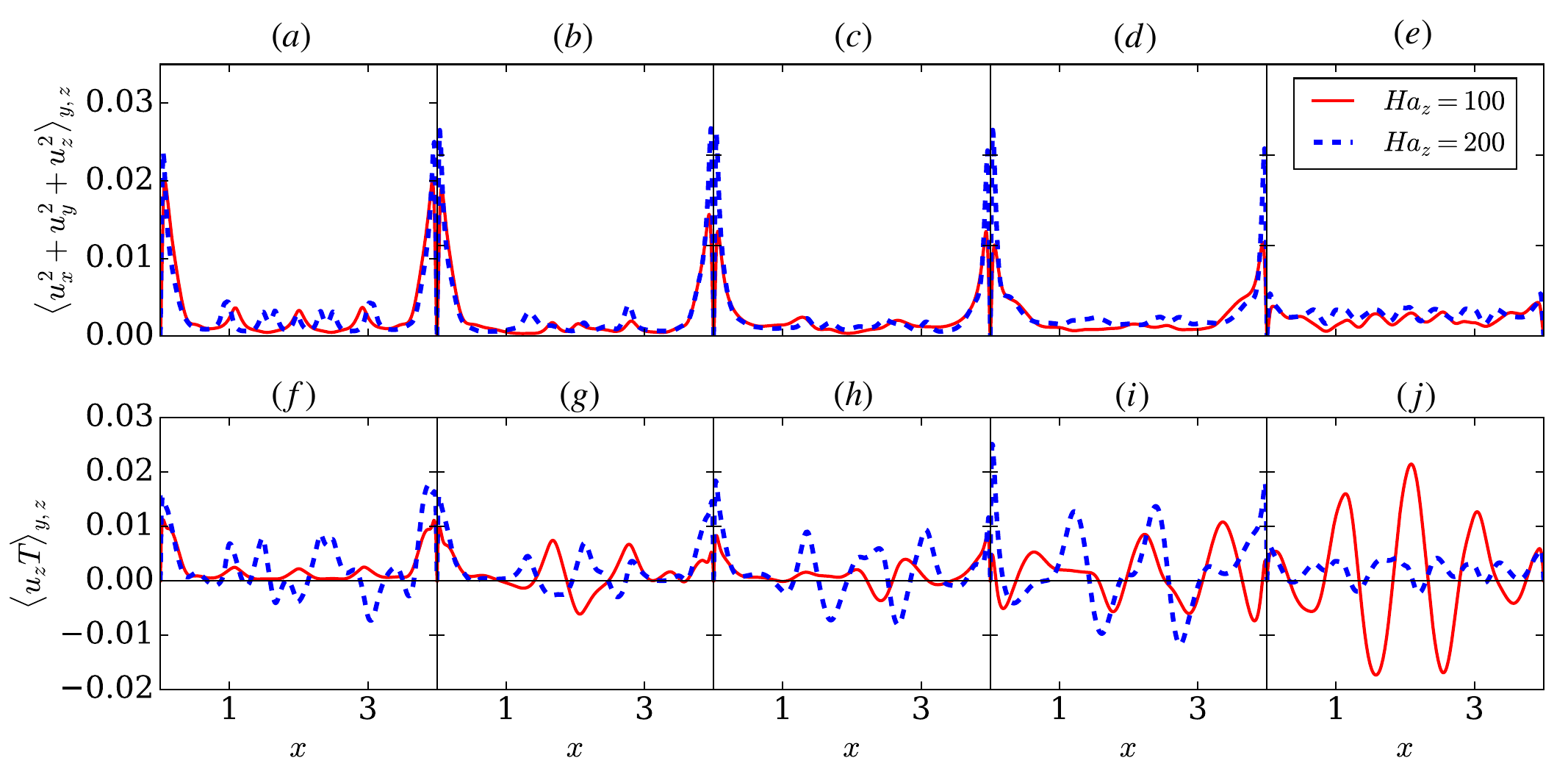}}
  \caption{Spatial distribution of kinetic energy $E$ and the convective heat flux $u_zT$. Plots of 2$\langle E \rangle_{y,z}=\langle u_x^2 + u_y^2 + u_z^2 \rangle_{y,z}$ for (a) $R=0$, (b) $R=0.3$, (c) $R=1$, (d) $R=2$, and (e) $R=3$. Plots of $\langle u_zT \rangle_{y,z}$ for (f) $R=0$, (g) $R=0.3$, (h) $R=1$, (i) $R=2$, and (j) $R=3$. In the above, $\langle \cdot \rangle_{y,z}$ represents averaging over $y$-$z$ plane.}
\label{fig:x_distribution}
\end{figure}

In figures~\ref{fig:x_distribution}(a)--(e), we plot twice the kinetic energy, averaged over the $y$-$z$ plane, versus $x$, which is the direction perpendicular to the horizontal magnetic field.
In the absence of horizontal magnetic field ($R=0$), the variation of $2\langle E \rangle_{y,z}$ versus $x$ is similar to that of $2\langle E \rangle_{x,z}$ versus $y$ due to the symmetry of the problem.
Again, there are two sharp peaks  corresponding to the wall modes close to $x=-2$ and $x=2$ for both $\Ha_z=100$ and $\Ha_z=200$.
However, unlike $\langle E \rangle_{x,z}$, the values of $\langle E \rangle_{y,z}$ at the peaks recede only marginally as $R$ is increased from 0 to 2.
As the wall modes extend fully into the bulk at $R=3$, the above peaks recede sharply with the value of $2\langle E \rangle_{y,z}$ at these peaks being close to its values at the peaks in the bulk.
It must be noted that $\langle E \rangle_{y,z}$ continues to fluctuate as it is varied with $x$ at $R=3$ and does not smoothen out unlike $\langle E \rangle_{x,z}$.

Figures~\ref{fig:x_distribution}(f--j) exhibit the variations of the local convective heat flux $u_zT$, averaged over the $y$-$z$ plane, along $x$.
The figures show that $\langle u_zT \rangle_{y,z}$ fluctuates between positive and negative values.
It is clear from the figures that for the $\Ha_z=100$ case, the spatial fluctuations increase with increasing $R$.
This is because as discussed before, the gradients along the $y$-direction are reduced as $R$ is increased. 
Therefore, for large $R$, if the gradients along $z$ are also small, any quantity averaged over $y$-$z$ plane will fluctuate between the quantity's two extremities. 
The case of $R=3$ for $\Ha_z=100$ consists of plumes from the opposite $y$=$\pm\Gamma$ walls merged with each other; therefore, the gradients of the velocity, temperature, and heat flux along the $z$-direction are small. 
Thus, $\langle u_zT \rangle_{y,z}$ exhibits strong fluctuations along the $x$-direction at $R=3$.

For $\Ha_z=200$ and $R \leq 2$, $\langle u_zT \rangle_{y,z}$ follows a similar trend in that its fluctuations increase with increasing $R$.
However, for $R=3$, its fluctuations get suppressed and $\langle u_zT \rangle_{y,z}$ is mostly positive.
Now, let us recall that unlike in the case of $\Ha_z=100$, the structures for the case of $\Ha_z=200$ are antisymmetric.
Thus, although the gradients along the $y$-direction are small, there are fluctuations along the $z$-direction for $\Ha_z=200$ due to the presence of upwelling and downwelling plumes on top of each other.
An averaging over the $y$-$z$ plane cancels out the opposing effects of the upwelling and downwelling plumes, thus resulting in the suppression of fluctuations of  $\langle u_zT \rangle_{y,z}$. 

\begin{figure}
  \centerline{\includegraphics[scale = 0.4]{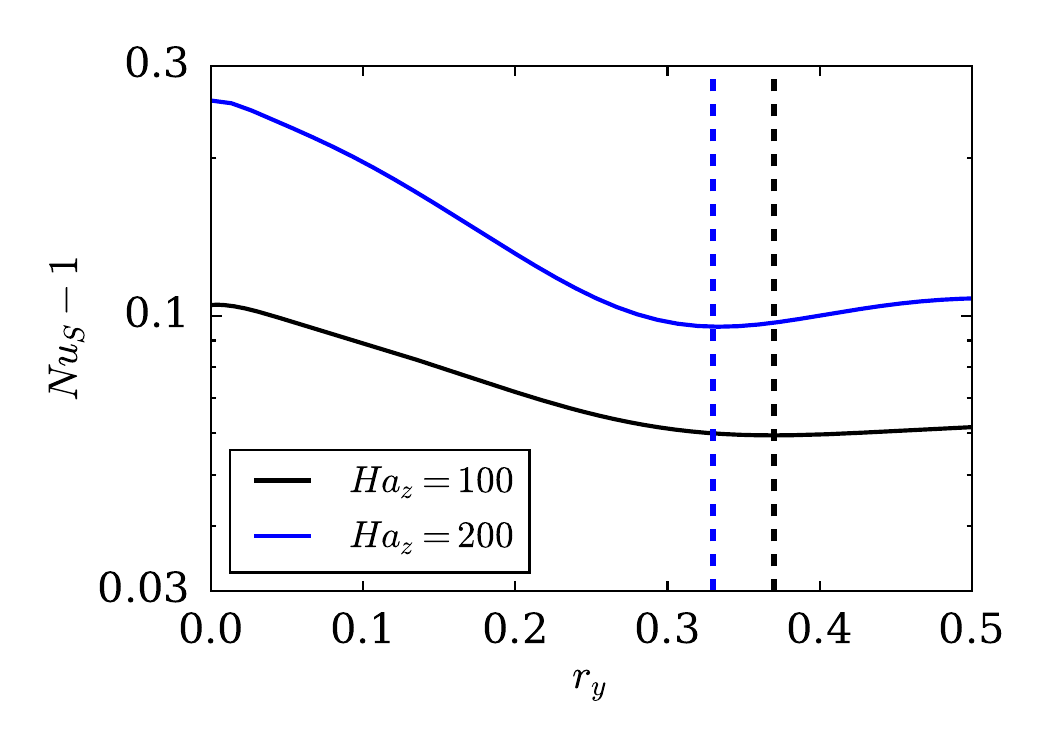}}
  \caption{Determination of the near-wall region for our analysis from the results of our DNS for $R=0$. Variation of the Nusselt number $\Nu_S$ averaged over successively smaller concentric volumes with the sidewall-normal distance $r_y$. The width of the near-wall region is given by the point of the first minima (represented as black dashed vertical line for $\Ha_z=100$ and blue dashed vertical line for $\Ha_z=200$) in the $\Nu_S$ profile  
 }
\label{fig:Near_wall}
\end{figure}

Finally, we analyze the contributions by the bulk and near-wall regions to the total kinetic energy and heat flux of the system. 
Towards this objective, we define the near-wall region as follows.
For $R=0$ and for both the Hartmann numbers, we determine $\Nu_S$, which is the Nusselt number averaged over successively smaller concentric volumes $S=\Gamma \times [r_y, \Gamma-r_y] \times 1$.  In this definition, $r_y$ is the normal distance from the $y$=$\pm\Gamma/2$ sidewalls.
We plot $\Nu_S$ versus $r_y$  in figure~\ref{fig:Near_wall}. 
The figure shows that $\Nu_S$ initially decreases with $r_y$ upto a point of local minima and then begins to increase with $r_y$. 
The distance between the point of the local minima and the sidewall is taken as the width $\delta_w$ of the near-wall region.
These widths are computed to be $\delta_w=0.37$ for $\Ha_z=100$ and $\delta_w=0.33$ for $\Ha_z=200$. Although the above values of $\delta_w$ were computed for $R=0$, these values will be assumed to hold for all $R$.

The total kinetic energy $\mathcal{E}$ and the convective heat flux $\mathcal{H}$ can be expressed as the sum of their bulk and near-wall contributions. Therefore,  
\begin{eqnarray}
    \mathcal{E} &=& \mathcal{E}_{bulk} + \mathcal{E}_{nw}, \label{eq:Tot_energy} \\
    \mathcal{H} &=& \mathcal{H}_{bulk} + \mathcal{H}_{nw}, \label{eq:Tot_heat_flux}
\end{eqnarray}
where $\mathcal{E}_{bulk}$ and $\mathcal{E}_{nw}$ are respectively the bulk and near-wall contributions to the total kinetic energy, and $\mathcal{H}_{bulk}$ and $\mathcal{H}_{nw}$ are respectively the bulk and near-wall contributions to the total heat flux.
The total and bulk contributions to these quantities are defined as
\begin{eqnarray}
 \mathcal{E} &=& \frac{1}{2}  \int_{-\Gamma/2}^{\Gamma/2} \int_{-\Gamma/2}^{\Gamma/2} \int_{-1/2}^{1/2} (u_x^2 + u_y^2 + u_z^2)~dz dy dx, \label{eq:Tot_energy_def}\\
    \mathcal{E}_{bulk} &=& \frac{1}{2}  \int_{-\Gamma/2+\delta_w}^{\Gamma/2-\delta_w} \int_{-\Gamma/2+\delta_w}^{\Gamma/2-\delta_w} \int_{-1/2}^{1/2} (u_x^2 + u_y^2 + u_z^2)~dz dy dx, 
    \label{eq:Bulk_energy_def}\\
    \mathcal{H} &=&  \int_{-\Gamma/2}^{\Gamma/2} \int_{-\Gamma/2}^{\Gamma/2} \int_{-1/2}^{1/2} u_zT~dz dy dx, \label{eq:Tot_heat_flux_def} \\
    \mathcal{H}_{bulk} &=&   \int_{-\Gamma/2+\delta_w}^{\Gamma/2-\delta_w} \int_{-\Gamma/2+\delta_w}^{\Gamma/2-\delta_w} \int_{-1/2}^{1/2} u_zT~dz dy dx \label{eq:Tot_heat_flux_def}. \\ 
\end{eqnarray}
We compute the relative strengths of the bulk and near-wall kinetic energies and heat fluxes using our numerical data and plot them versus $R$ in figures~\ref{fig:Near_wall}(a)--(d).
For small values of $R$, the bulk contribution to the total kinetic energy and heat flux is very small (less than 10\%). 
This is expected because convection is completely suppressed in the bulk at small $R$. 
It is clear from the figures that as $R$ increases, the bulk contributions to the total kinetic energy and heat flux increase. 
This is due to the fact that the wall modes extend further into bulk as $R$ increases. 
In fact, for $R>2$, the bulk and sidewall contributions become comparable to each other because the wall modes at such high values of $R$ extend fully into the bulk.
It is also interesting to note that for $\Ha_z=100$, the bulk contribution to the heat flux decreases as $R$ is increased from 2 to 3. 
The reason for this anomalous behaviour is yet to be understood.

\begin{figure}
  \centerline{\includegraphics[scale = 0.325]{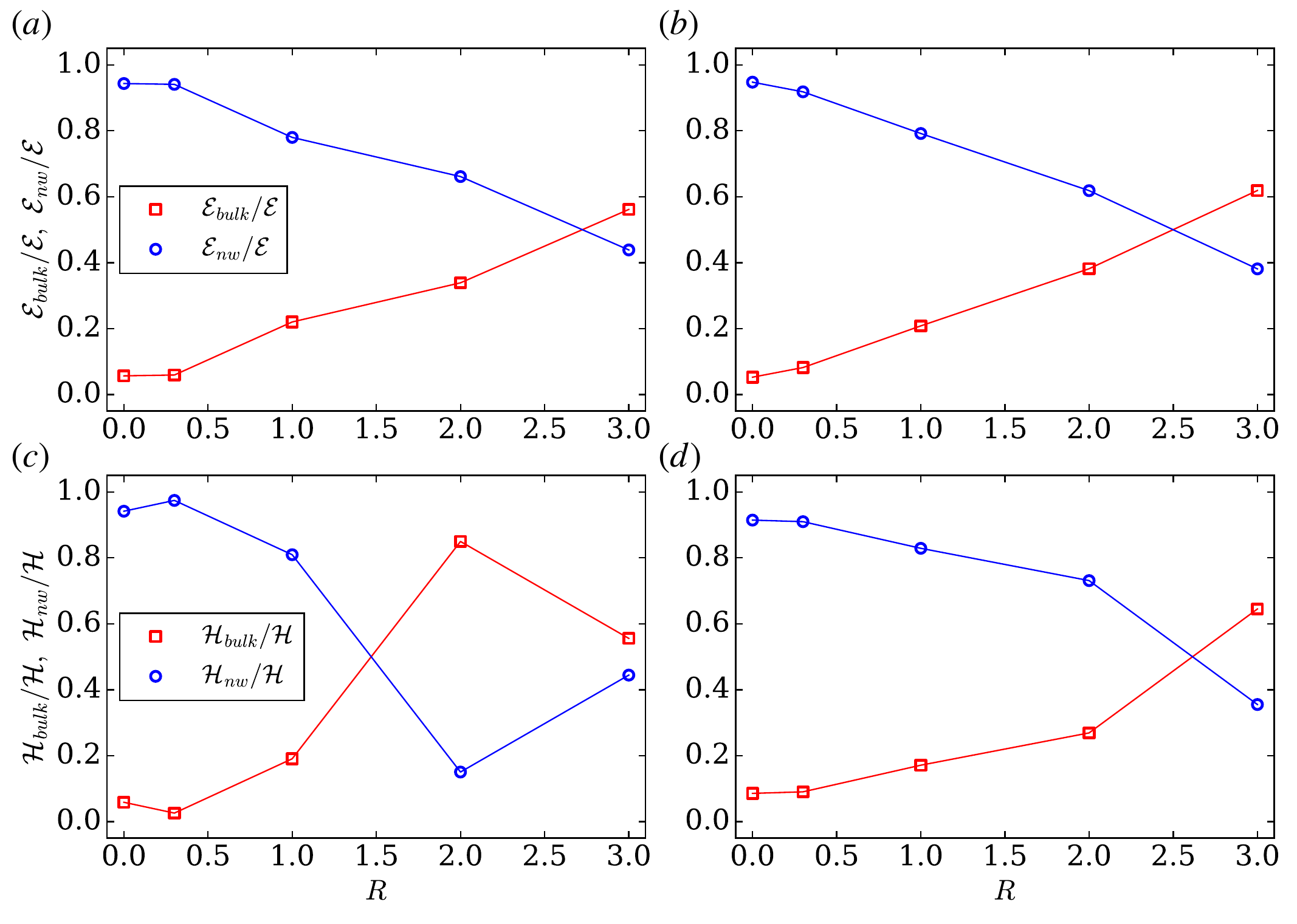}}
  \caption{Results of direct numerical simulations: Relative strengths of the bulk and boundary layer contributions to the total kinetic energy $\mathcal{E}$ for (a) $\Ha_z=100$ and (b) $\Ha_z=200$, and to the total heat flux $\mathcal{F}$ for (c) $\Ha_z=100$ and (d) $\Ha_z=200$. The bulk contributions to kinetic energy and heat flux increase with increasing $R$. 
 }
\label{fig:Contributions}
\end{figure}

\subsubsection{Effects of initial conditions} \label{sec:Init_cond}
The direct numerical simulations of magnetoconvection for $R=3$ resulted in different structural arrangements of the convection plumes for $\Ha_z=100$ and $\Ha_z=200$. 
On one hand, we obtained a symmetric arrangement of the plumes adjacent to the opposite walls merging with each other for $\Ha_z=100$.
On the other hand, an antisymmetric arrangement of the plumes was obtained for $\Ha_z=200$ with the upwelling and downwelling plumes on top of each other.
In this subsection, we will explore the sensitivity of the above results to initial conditions.

We conduct two more direct numerical simulations of magnetoconvection for $R=3$; one with $\Ha_z=100$, $\Ray=10^5$, and the other with $\Ha_z=200$, $\Ray = 4 \times 10^5$. 
However, we change the initial conditions in this set of simulations as follows. 
The results of the old simulation of $\Ha_z=200$ are taken as the initial condition for the new simulation of $\Ha_z=100$.
In the same way, the results of the old simulation of $\Ha_z=100$ are taken as the initial condition for the new simulation of $\Ha_z=200$. 
Both these simulations were allowed to run for 20 free-fall time units after reaching the steady state. 
Henceforth, we refer to the old set of simulations as IC1 and the new set of simulations of $R=3$ as IC2.

\begin{figure}
  \centerline{\includegraphics[scale = 0.25]{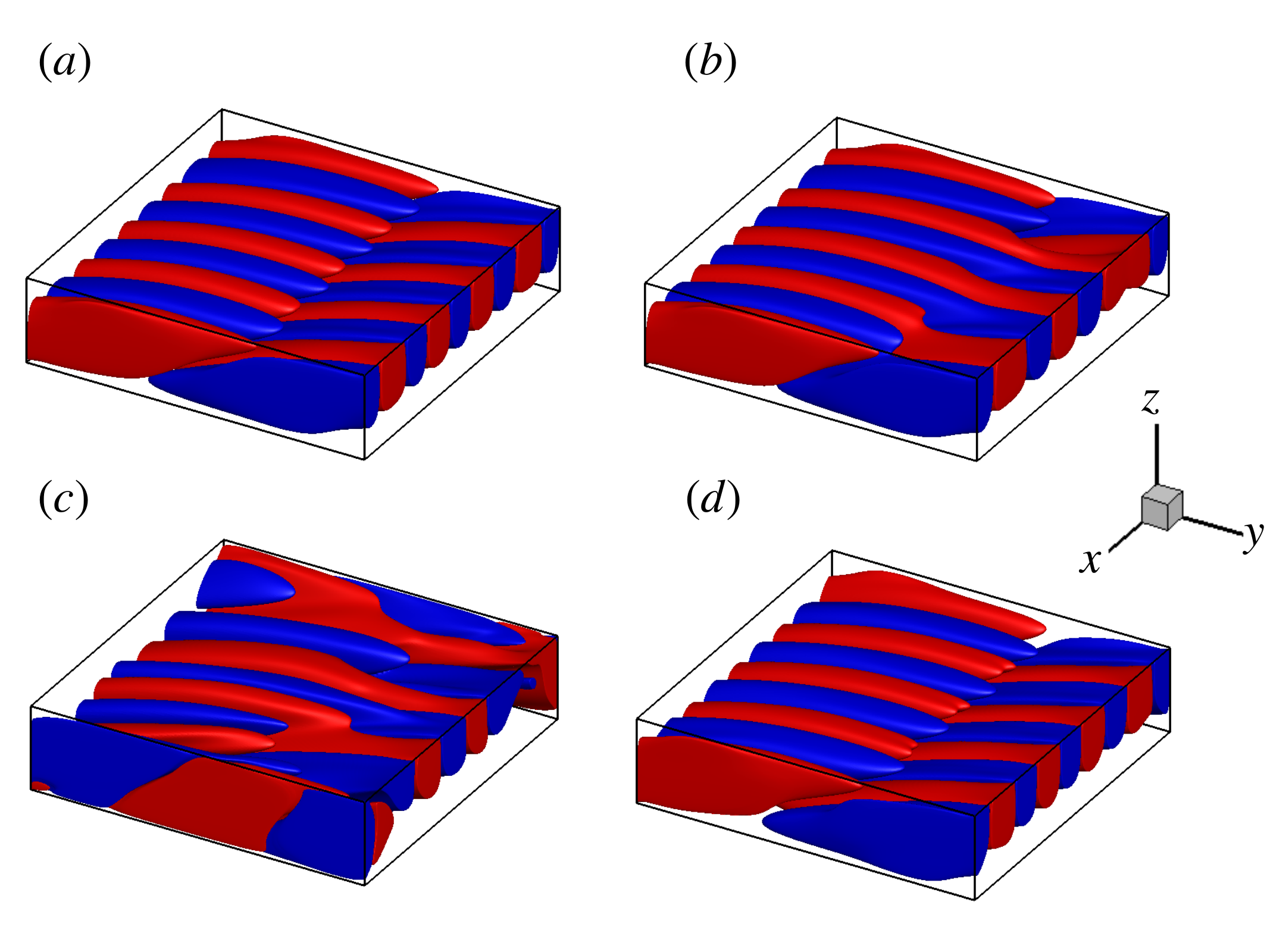}}
  \caption{Results of direct numerical simulations with new sets of initial conditions (IC2): Isosurfaces of $u_z=0.01$ (red) and $u_z=-0.01$ (blue) for $R=3$ and (a) $Ha_z=100$ and (c) $\Ha_z=200$. For comparison, the vertical velocity isosurfaces using results of the simulations of $R=3$ with the old initial conditions are shown for (b) $\Ha_z=100$ and (d) $\Ha_z=200$}
\label{fig:Wall_modes_isosurface_diff}
\end{figure}

We examine the structure of the wall-modes for the new set of simulations by plotting the vertical velocity isosurfaces for $u_z = \pm 0.01$ in figures~\ref{fig:Wall_modes_isosurface_diff}(a) and (c). 
The figures show that the simulation of $\Ha_z=100$ with the new initial conditions yields antisymmetric arrangement structures with upwelling and downwelling plumes on top of each other. 
This is unlike the result of the simulation with old initial conditions which resulted in a symmetric arrangement of structures with merged plumes.
Similarly, the simulation of $\Ha_z=200$ with the new initial conditions yield symmetric structures with merged plumes, unlike the case with old initial conditions which resulted in antisymmetric arrangement of structures.
Our results indicate that that the solution of magnetoconvection with $R=3$ for $\Ha_z=100$ and 200 is non-unique, and the arrangement of the wall modes depends on the initial conditions.

\begin{figure}
  \centerline{\includegraphics[scale = 0.35]{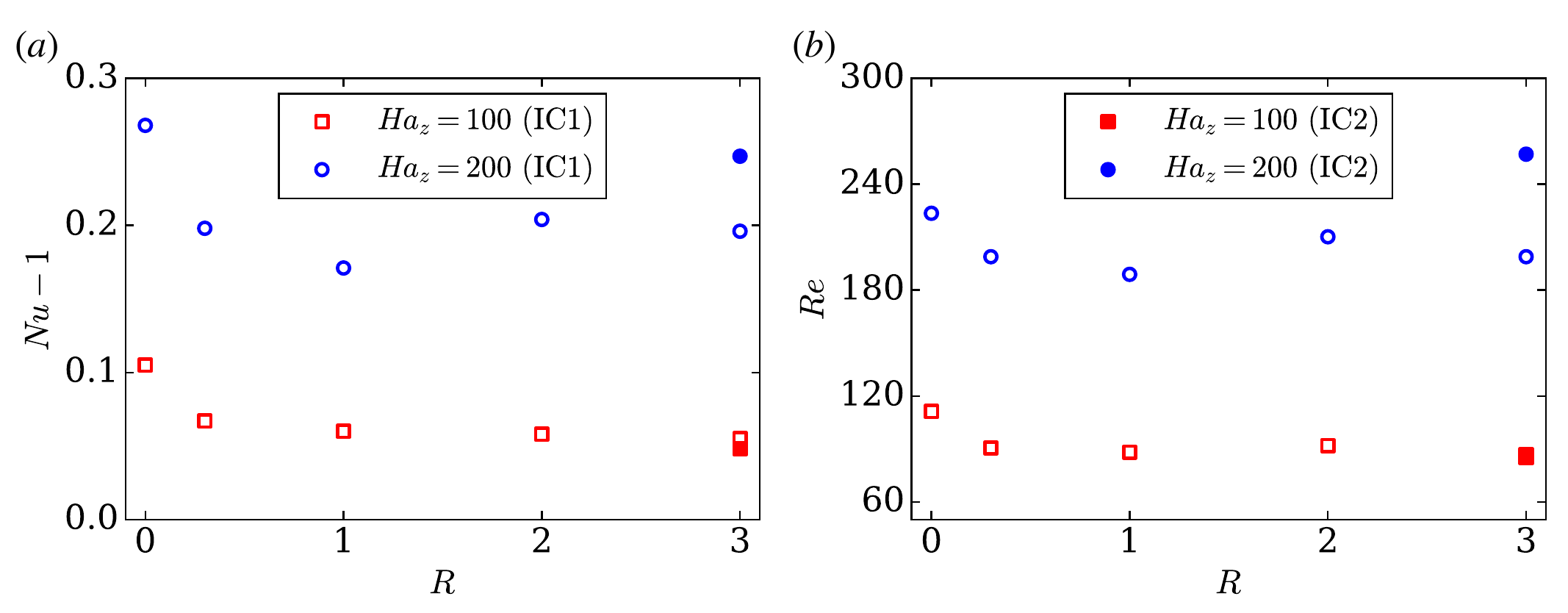}}
  \caption{Results of direct numerical simulations including those using the new initial conditions (filled markers) along with the old initial conditions (unfilled markers): Plots of (a) the nondimensional convective heat flux $\Nu-1$ and (b) the Reynolds number $\Rey$ versus $R$. The heat and momentum transport for the case of $\Ha_z=200$, $R=3$ is significantly more for the new initial condition (IC2) compared to the old initial condition (IC1)}
\label{fig:NuRe_diffInit}
\end{figure}
We will now explore the impact of the arrangement of the plumes on the global heat and momentum transport. In figures~\ref{fig:NuRe_diffInit}(a,b), we replot the computed values of $\Nu-1$ and $\Rey$ for our runs with the old initial conditions and also add the plots of these quantities computed using the results of our simulations with new initial conditions. 
The figures show that for $\Ha_z=200$, the Reynolds and Nusselt numbers computed using the solutions of simulations with new initial conditions are significantly higher than those corresponding to the old initial conditions.
On the other hand, for $\Ha_z=100$, the Reynolds and Nusselt numbers computed using the solutions of simulations with new initial conditions are lower, albeit marginally, than those corresponding to the old initial conditions.
The above observations clearly indicate that the solutions that are symmetric consisting of merged plumes have higher heat and momentum transport.
This is because merged plumes give rise to larger convection rolls which, in turn, result in more efficient transport of heat and momentum.
It is worth noting that for $\Ha_z=200$, the increase of Reynolds number at $R=3$ due to the merged plumes is so significant that its value is larger than for $R=0$.

Our results indicate that the non-unique nature of the solutions for $R=3$ has a profound effect on the heat and momentum transport, especially for $\Ha_z=200$. We conclude in the next section.

\section{Summary and conclusions} \label{sec:Conclusions}
In this paper, we systematically examined the effects of strong inclined magnetic fields on wall-attached magnetoconvection using a combination of linear stability analysis and direct numerical simulations.
The linear stability analysis was conducted for lower Hartmann numbers and aspect-ratio box whereas the DNS were conducted for higher Hartmann numbers and aspect ratio.
The ratio $R=B_y/B_z$ of the outer imposed horizontal to vertical magnetic field was varied from 0 to 3.

The linear stability analysis revealed that the critical Rayleigh number varies non-monotonically with the relative strength of the horizontal magnetic field.
The plumes at the onset of instability get elongated along the direction the resultant magnetic field and extend more into the bulk as $R$ is increased.
At sufficiently high $R$, the plumes extend fully into the bulk. The linear stability analysis further revealed the existence of non-unique solutions at the onset at low $R$. 
One eigensolution corresponds to a symmetric arrangement of hot or cold plumes adjacent to the opposite sidewalls, whereas the other eigensolution corresponds to an antisymmetric arrangement of hot or cold plumes.
For the symmetric solution, when the opposite plumes extend sufficiently into the bulk at a large $R$, they merge into a single large plume.
Below this critical value of $R=R_c$, the symmetric and antisymmetric eigensolutions overlap; however, at $R>R_c$, these solutions start to diverge with the symmetric eigensolution being more unstable than the antisymmetric solution.
The critical Rayleigh number increases with $R$ for $R<R_c$ and decreases with $R$ for $R>R_c$.

The direct numerical simulations of the fully nonlinear regime reinforced the observation from the linear stability analysis that the wall modes get elongated along the direction of the resultant magnetic field and extend further into the bulk as $R$ is increased.
The heat and momentum transport was observed to decrease with $R$ for $R<1$ but not show any visible trend for $R>1$.
The fluctuations of the local heat flux and kinetic energy along the direction of the horizontal magnetic field get suppressed as $R$ is increased.
Since the wall modes extend more into the bulk as $R$ is increased, the relative contribution to the total heat transport and kinetic energy by the near-wall regions decrease with an increase of $R$.

The analysis from direct numerical simulations further revealed that at least for $R=3$, the solutions are dependent on the initial conditions. 
The solutions for $R=3$ can either comprise of an antisymmetric arrangement of upwelling and downwelling plumes on top of each other, or a symmetric arrangement of merged upwelling or downwelling plumes.
The solution with merged plumes corresponds to higher heat and momentum transport because the merged plumes give rise to larger convection rolls and hence more efficient heat transfer.

Our present work provides important insights into the dynamics of wall-attached convection under inclined magnetic fields which will be a typical situation in view to most applications. Our work may find applications in several industrial  flows such as cooling blankets in fusion reactors.  Although we worked on a small set of parameters, we expect our results to hold for higher Rayleigh numbers as well. In the future, we plan to conduct a similar analysis for fluids at different Prandtl numbers.

\section*{Acknowledgements}
The authors thank M. K. Verma and M. Brynjell-Rahkola for useful discussions. The authors acknowledge the computing time provided by Leibniz
Supercomputing Center, Garching, Germany, under the project ‘pn49ma’. The authors further acknowledge the Computing Center of Technische Universit{\"a}t Ilmenau for the resources provided to them for the linear stability computations as well as for the postprocessing and visualization of their simulation data.

\section*{Funding}
The work of S.B. is sponsored by a Postdoctoral Fellowship of the Alexander von Humboldt Foundation, Germany.

\section*{Declaration of interests}
The authors report no conflict of interest.

\section*{Author ORCIDs}
Shashwat Bhattacharya https://orcid.org/0000-0001-7462-7680

Thomas Boeck https://orcid.org/0000-0002-0814-7432

Dmitry Krasnov https://orcid.org/0000-0002-8339-7749

J{\"o}rg Schumacher https://orcid.org/0000-0002-1359-4536

\bibliographystyle{jfm}

\begin{thebibliography}{44}
	\expandafter\ifx\csname natexlab\endcsname\relax\def\natexlab#1{#1}\fi
	\def\au#1{#1} \def\ed#1{#1} \def\yr#1{#1}\def\at#1{#1}\def\jt#1{\textit{#1}}
	\def\bt#1{#1}\def\bvol#1{\textbf{#1}} \def\vol#1{#1} \def\pg#1{#1}
	\def\publ#1{#1}\def\arxiv#1{#1}\def\org#1{#1}\def\st#1{\textit{#1}}
	
	\bibitem[Akhmedagaev {\em et~al.\/}(2020{\natexlab{{\em a\/}}})Akhmedagaev,
	Zikanov, Krasnov \& Schumacher]{Akhmedagaev:MHD2020}
	{\sc \au{Akhmedagaev, R.}, \au{Zikanov, O.}, \au{Krasnov, D.} \&
		\au{Schumacher, J.}} \yr{2020{\natexlab{{\em a\/}}}}
	\at{{Rayleigh-B{\'e}nard convection in strong vertical magnetic field: flow
			structure and verification of numerical method}}.  \jt{Magnetohydrodynamics}
	\bvol{56},  \pg{157--166}.
	
	\bibitem[Akhmedagaev {\em et~al.\/}(2020{\natexlab{{\em b\/}}})Akhmedagaev,
	Zikanov, Krasnov \& Schumacher]{Akhmedagaev:JFM2020}
	{\sc \au{Akhmedagaev, R.}, \au{Zikanov, O.}, \au{Krasnov, D.} \&
		\au{Schumacher, J.}} \yr{2020{\natexlab{{\em b\/}}}}  \at{{Turbulent
			Rayleigh-B{\'e}nard convection in a strong vertical magnetic field}}.  \jt{J.
		Fluid Mech.}  \bvol{895},  \pg{R4}.
	
	\bibitem[Aurnou \& Olson(2001)]{Aurnou:JFM2001}
	{\sc \au{Aurnou, J.~M.} \& \au{Olson, P.~L.}} \yr{2001}  \at{{Experiments on
			Rayleigh–B{\'e}nard convection, magnetoconvection and rotating
			magnetoconvection in liquid gallium}}.  \jt{J. Fluid Mech.}  \bvol{430},
	\pg{283--307}.
	
	\bibitem[Bhattacharya {\em et~al.\/}(2023)Bhattacharya, Boeck, Krasnov \&
	Schumacher]{Bhattacharya:JFM2023}
	{\sc \au{Bhattacharya, S.}, \au{Boeck, T.}, \au{Krasnov, D.} \& \au{Schumacher,
			J.}} \yr{2023}  \at{{Effects of strong fringing magnetic fields on turbulent
			thermal convection}}.  \jt{J. Fluid Mech.}  \bvol{964},  \pg{A31}.
	
	\bibitem[Burr \& M{\"u}ller(2001)]{Burr:PF2001}
	{\sc \au{Burr, U.} \& \au{M{\"u}ller, U.}} \yr{2001}
	\at{{Rayleigh–B{\"e}nard convection in liquid metal layers under the
			influence of a vertical magnetic field.}}  \jt{Phys. Fluids}  \bvol{13},
	\pg{3247--3257}.
	
	\bibitem[Burr \& M{\"u}ller(2002)]{Burr:JFM2002}
	{\sc \au{Burr, U.} \& \au{M{\"u}ller, U.}} \yr{2002}
	\at{{Rayleigh{\textendash}B{\'e}nard convection in liquid metal layers under
			the influence of a horizontal magnetic field}}.  \jt{J. Fluid Mech.}
	\bvol{453},  \pg{345--369}.
	
	\bibitem[Busse(2008)]{Busse:PF2008}
	{\sc \au{Busse, F.~H.}} \yr{2008}  \at{Asymptotic theory of wall-attached
		convection in a horizontal fluid layer with a vertical magnetic field}.
	\jt{Phys. Fluids}  \bvol{20}~(2),  \pg{024102}.
	
	\bibitem[Busse \& Clever(1983)]{Busse:JTAM1983}
	{\sc \au{Busse, F.~H.} \& \au{Clever, R.~M.}} \yr{1983}  \at{{Stability of
			convection rolls in the presence of a horizontal magnetic field}}.  \jt{J.
		Theor. Appl. Mech.}  \bvol{2},  \pg{495--502}.
	
	\bibitem[Chandrasekhar(1981)]{Chandrasekhar:book:Instability}
	{\sc \au{Chandrasekhar, S.}} \yr{1981} {\em {Hydrodynamic and Hydromagnetic
			Stability}\/}.  \publ{Oxford: Dover publications}.
	
	\bibitem[Chill{\`a} \& Schumacher(2012)]{Chilla:EPJE2012}
	{\sc \au{Chill{\`a}, F.} \& \au{Schumacher, J.}} \yr{2012}  \at{{New
			perspectives in turbulent Rayleigh-B{\'e}nard convection}}.  \jt{Eur. Phys.
		J. E}  \bvol{35}~(7),  \pg{58}.
	
	\bibitem[Cioni {\em et~al.\/}(2000)Cioni, Chaumat \& Sommeria]{Cioni:PRL2000}
	{\sc \au{Cioni, S}, \au{Chaumat, S} \& \au{Sommeria, J}} \yr{2000}  \at{{Effect
			of a vertical magnetic field on turbulent Rayleigh-B{\'e}nard convection}}.
	\jt{Phys. Rev. Lett.}  \bvol{62}~(4),  \pg{R4520--R4523}.
	
	\bibitem[Davidson(1999)]{Davidson:ARFM1999}
	{\sc \au{Davidson, Peter~A.}} \yr{1999}  \at{{Magnetohydrodynamics in materials
			processing}}.  \jt{Annu. Rev. Fluid Mech.}  \bvol{31}~(1),  \pg{273--300}.
	
	\bibitem[Davidson(2017)]{Davidson:book:MHD}
	{\sc \au{Davidson, Peter~A.}} \yr{2017} {\em {An Introduction to
			Magnetohydrodynamics}\/}, 2nd edn.  \publ{Cambridge: Cambridge University
		Press}.
	
	\bibitem[Fauve {\em et~al.\/}(1981)Fauve, Laroche \& Libchaber]{Fauve:JPL1981}
	{\sc \au{Fauve, S.}, \au{Laroche, C.} \& \au{Libchaber, A.}} \yr{1981}
	\at{{Effect of a horizontal magnetic field on convective instabilities in
			mercury}}.  \jt{J. Physique Lett.}  \bvol{42}~(21),  \pg{L455}.
	
	\bibitem[Grannan {\em et~al.\/}(2022)Grannan, Cheng, Aggarwal, Hawkins, Xu,
	Horn, S{\'a}nchez-{\'A}lvarez \& Aurnou]{Grannan:JFM2022}
	{\sc \au{Grannan, A.~M.}, \au{Cheng, J.~S.}, \au{Aggarwal, A.}, \au{Hawkins,
			E.~K.}, \au{Xu, Y.}, \au{Horn, S.}, \au{S{\'a}nchez-{\'A}lvarez, J.} \&
		\au{Aurnou, J.~M.}} \yr{2022}  \at{{Experimental pub crawl from
			Rayleigh-B{\'e}nard to magnetostrophic convection}}.  \jt{J. Fluid Mech.}
	\bvol{939},  \pg{R1}.
	
	\bibitem[Houchens {\em et~al.\/}(2002)Houchens, Witkowski \&
	Walker]{Houchens:JFM2002}
	{\sc \au{Houchens, B.~C.}, \au{Witkowski, L.~Martin} \& \au{Walker, J.~S.}}
	\yr{2002}  \at{Rayleigh–bénard instability in a vertical cylinder with a
		vertical magnetic field}.  \jt{J. Fluid Mech.}  \bvol{469},  \pg{189–207}.
	
	\bibitem[Hurlburt {\em et~al.\/}(1996)Hurlburt, Matthews \&
	Proctor]{HurlBurt:APJ1996}
	{\sc \au{Hurlburt, N.}, \au{Matthews, P.} \& \au{Proctor, M.}} \yr{1996}
	\at{{Nonlinear compressible convection in oblique magnetic fields}}.  \jt{Ap.
		J.}  \bvol{457},  \pg{933}.
	
	\bibitem[Kelley \& Sadoway(2014)]{Kelley:PF2014}
	{\sc \au{Kelley, D.} \& \au{Sadoway, D.~R.}} \yr{2014}  \at{{Mixing in a liquid
			metal electrode}}.  \jt{Phys. Fluids}  \bvol{26},  \pg{057102}.
	
	\bibitem[Kelley \& Weier(2018)]{Kelley:AMR2018}
	{\sc \au{Kelley, D.} \& \au{Weier, T.}} \yr{2018}  \at{{Fluid Mechanics of
			Liquid Metal Batteries}}.  \jt{Appl. Mech. Rev.}  \bvol{70},  \pg{020801}.
	
	\bibitem[King \& Aurnou(2015)]{King:PNAS2015}
	{\sc \au{King, E.~M.} \& \au{Aurnou, J.~M.}} \yr{2015}  \at{{Magnetostrophic
			balance as the optimal state for turbulent magnetoconvection}}.  \jt{Proc.
		Natl. Acad. Sci. U.S.A.}  \bvol{112},  \pg{990--994}.
	
	\bibitem[Krasnov {\em et~al.\/}(2011)Krasnov, Zikanov \& Boeck]{Krasnov:CF2011}
	{\sc \au{Krasnov, D.}, \au{Zikanov, O.} \& \au{Boeck, T.}} \yr{2011}
	\at{Comparative study of finite difference approaches in simulation of
		magnetohydrodynamic turbulence at low magnetic reynolds number}.  \jt{Comput.
		Fluids}  \bvol{50}~(1),  \pg{46 -- 59}.
	
	\bibitem[Liu {\em et~al.\/}(2018)Liu, Krasnov \& Schumacher]{Liu:JFM2018}
	{\sc \au{Liu, W.}, \au{Krasnov, D.} \& \au{Schumacher, J.}} \yr{2018}
	\at{{Wall modes in magnetoconvection at high Hartmann numbers}}.  \jt{J.
		Fluid Mech.}  \bvol{849},  \pg{R2}.
	
	\bibitem[Lohse \& Xia(2010)]{Lohse:ARFM2010}
	{\sc \au{Lohse, Detlef} \& \au{Xia, Ke-Qing}} \yr{2010}  \at{{Small-scale
			properties of turbulent Rayleigh{\textendash}B{\'e}nard convection}}.
	\jt{Annu. Rev. Fluid Mech.}  \bvol{42}~(1),  \pg{335--364}.
	
	\bibitem[Lyubimov {\em et~al.\/}(2010)Lyubimov, Burnysheva, Benhadid, Lyubimova
	\& Henry]{Lyubimov:JFM2010}
	{\sc \au{Lyubimov, D.~V.}, \au{Burnysheva, A.~V.}, \au{Benhadid, H.},
		\au{Lyubimova, T.~P.} \& \au{Henry, D.}} \yr{2010}  \at{{Rotating magnetic
			field effect on convection and its stability in a horizontal cylinder
			subjected to a longitudinal temperature gradient}}.  \jt{J. Fluid Mech.}
	\bvol{664},  \pg{108--137}.
	
	\bibitem[McCormack {\em et~al.\/}(2023)McCormack, Teimurazov, Shishkina \&
	Linkmann]{mccormack2023wall}
	{\sc \au{McCormack, Matthew}, \au{Teimurazov, Andrei}, \au{Shishkina, Olga} \&
		\au{Linkmann, Moritz}} \yr{2023} Wall mode dynamics and transition to chaos
	in magnetoconvection with a vertical magnetic field,  \arxiv{arXiv:
		2308.15165}.
	
	\bibitem[Mistrangelo {\em et~al.\/}(2020)Mistrangelo, B{\"u}hler \&
	Kl{\"u}ber]{Mistrangelo:FED2020}
	{\sc \au{Mistrangelo, C.}, \au{B{\"u}hler, L.} \& \au{Kl{\"u}ber, V.}}
	\yr{2020}  \at{{Three-dimensional magneto convective flows in geometries
			relevant for DCLL blankets}}.  \jt{Fusion Eng. Des.}  \bvol{159},
	\pg{111686}.
	
	\bibitem[Mistrangelo {\em et~al.\/}(2021)Mistrangelo, B{\"u}hler, Smolentsev,
	Kl{\"u}ber, Maione \& Aubert]{Mistrangelo:FED2021}
	{\sc \au{Mistrangelo, C.}, \au{B{\"u}hler, L.}, \au{Smolentsev, S.},
		\au{Kl{\"u}ber, V.}, \au{Maione, I.} \& \au{Aubert, J.}} \yr{2021}  \at{{MHD
			flow in liquid metal blankets: Major design issues, MHD guidelines and
			numerical analysis}}.  \jt{Fusion Eng. Des.}  \bvol{173},  \pg{112795}.
	
	\bibitem[Nakagawa(1957)]{Nakagawa:PRSL1957}
	{\sc \au{Nakagawa, Y}} \yr{1957}  \at{{Experiments on the inhibition of thermal
			convection by a magnetic field}}.  \jt{Proc. R. Soc. Lond.}  \bvol{240},
	\pg{108--113}.
	
	\bibitem[Nicoski {\em et~al.\/}(2022)Nicoski, Yan \& Calkins]{Nicoski:PRF2022}
	{\sc \au{Nicoski, J.~A.}, \au{Yan, M.} \& \au{Calkins, M.~A.}} \yr{2022}
	\at{{Quasistatic magnetoconvection with a tilted magnetic field}}.  \jt{Phys.
		Rev. Fluids}  \bvol{7},  \pg{043504}.
	
	\bibitem[Peyret(2002)]{Peyret:book}
	{\sc \au{Peyret, R}} \yr{2002} {\em {Spectral Methods for Incompressible
			Viscous Flows}\/}.  \publ{New York: Springer}.
	
	\bibitem[Priede {\em et~al.\/}(2010)Priede, Aleksandrova \&
	Molokov]{Priede:JFM2010}
	{\sc \au{Priede, J{\=a}nis}, \au{Aleksandrova, Svetlana} \& \au{Molokov,
			Sergei}} \yr{2010}  \at{Linear stability of {H}unt's flow}.  \jt{J. Fluid
		Mech.}  \bvol{649},  \pg{115--134}.
	
	\bibitem[Roberts(1967)]{Roberts:book}
	{\sc \au{Roberts, P.~H.}} \yr{1967} {\em {An introduction to
			magnetohydrodynamics}\/}.  \publ{London: Longmans}.
	
	\bibitem[Shen \& Zikanov(2016)]{Shen:TCFD2016}
	{\sc \au{Shen, Y.} \& \au{Zikanov, O.}} \yr{2016}  \at{{Thermal convection in a
			liquid metal battery}}.  \jt{Theor. Comput. Fluid Dyn.}  \bvol{30},
	\pg{275--294}.
	
	\bibitem[Tasaka {\em et~al.\/}(2016)Tasaka, Igaki, Yanagisawa, Vogt, Z{\"u}rner
	\& Eckert]{Tasaka:PRE2016}
	{\sc \au{Tasaka, Y.}, \au{Igaki, K.}, \au{Yanagisawa, T.}, \au{Vogt, T.},
		\au{Z{\"u}rner, T.} \& \au{Eckert, S.}} \yr{2016}  \at{{Regular flow
			reversals in Rayleigh-B{\'e}nard convection in a horizontal magnetic field}}.
	\jt{Phys. Rev. E}  \bvol{93},  \pg{043109}.
	
	\bibitem[Teimurazov {\em et~al.\/}(2023)Teimurazov, McCormack, Linkmann \&
	Shishkina]{teimurazov2023unifying}
	{\sc \au{Teimurazov, Andrei}, \au{McCormack, Matthew}, \au{Linkmann, Moritz} \&
		\au{Shishkina, Olga}} \yr{2023} Unifying heat transport model for the
	transition between buoyancy-dominated and lorentz-force-dominated regimes in
	quasistatic magnetoconvection,  \arxiv{arXiv: 2308.01748}.
	
	\bibitem[{The MathWorks Inc.}(2022)]{MATLAB}
	{\sc \au{{The MathWorks Inc.}}} \yr{2022} Matlab version: 9.13.0 (r2022b).
	
	\bibitem[Verma(2018)]{Verma:book:BDF}
	{\sc \au{Verma, Mahendra~K.}} \yr{2018} {\em Physics of Buoyant Flows: From
		Instabilities to Turbulence\/}.  \publ{Singapore: World Scientific}.
	
	\bibitem[Verma(2019)]{Verma:ET}
	{\sc \au{Verma, M.~K.}} \yr{2019} {\em Energy trasnfers in Fluid Flows:
		Multiscale and Spectral Perspectives\/}.  \publ{Cambridge: Cambridge
		University Press}.
	
	\bibitem[Vogt {\em et~al.\/}(2018)Vogt, Ishimi, Yanagisawa, Tasaka, Sakuraba \&
	Eckert]{Vogt:PRF2018}
	{\sc \au{Vogt, T.}, \au{Ishimi, W.}, \au{Yanagisawa, T.}, \au{Tasaka, Y.},
		\au{Sakuraba, A.} \& \au{Eckert, S.}} \yr{2018}  \at{{Transition between
			quasi-two-dimensional and three-dimensional Rayleigh-B{\'e}nard convection in
			a horizontal magnetic field}}.  \jt{Phys. Rev. Fluids}  \bvol{3},
	\pg{013503}.
	
	\bibitem[Vogt {\em et~al.\/}(2021)Vogt, Yang, Schindler \&
	Eckert]{Vogt:JFM2021}
	{\sc \au{Vogt, T.}, \au{Yang, J.-C.}, \au{Schindler, F.} \& \au{Eckert, S.}}
	\yr{2021}  \at{{Free-fall velocities and heat transport enhancement in liquid
			metal magneto-convection}}.  \jt{J. Fluid Mech.}  \bvol{915},  \pg{A68}.
	
	\bibitem[Weiss \& Proctor(2014)]{Weiss:book}
	{\sc \au{Weiss, N.~O.} \& \au{Proctor, M. R.~E.}} \yr{2014} {\em
		{Magnetoconvection}\/}.  \publ{Cambridge: Cambridge University Press}.
	
	\bibitem[Yan {\em et~al.\/}(2019)Yan, Calkins, Maffei, Julien, Tobias \&
	Marti]{Yan:JFM2019}
	{\sc \au{Yan, M.}, \au{Calkins, M.}, \au{Maffei, A.}, \au{Julien, K.},
		\au{Tobias, S.} \& \au{Marti, P.}} \yr{2019}  \at{{Heat transfer and flow
			regimes in quasi-static magnetoconvection with a vertical magnetic field}}.
	\jt{J. Fluid Mech.}  \bvol{877},  \pg{1186--1206}.
	
	\bibitem[Yanagisawa {\em et~al.\/}(2013)Yanagisawa, Hamano, Miyagoshi,
	Yamagishi, Tasaka \& Takeda]{Yanagisawa:PRE2013}
	{\sc \au{Yanagisawa, T.}, \au{Hamano, Y.}, \au{Miyagoshi, T.}, \au{Yamagishi,
			Y.}, \au{Tasaka, Y.} \& \au{Takeda, Y.}} \yr{2013}  \at{{Convection patterns
			in a liquid metal under an imposed horizontal magnetic field}}.  \jt{Phys.
		Rev. E}  \bvol{88}~(6),  \pg{063020}.
	
	\bibitem[Z\"urner {\em et~al.\/}(2020)Z\"urner, Schindler, Vogt, Eckert \&
	Schumacher]{Zuerner:JFM2020}
	{\sc \au{Z\"urner, T.}, \au{Schindler, F.}, \au{Vogt, T.}, \au{Eckert, S.} \&
		\au{Schumacher, J.}} \yr{2020}  \at{{Flow regimes of Rayleigh-B{\'e}nard
			convection in a vertical magnetic field}}.  \jt{J. Fluid Mech.}  \bvol{894},
	\pg{A21}.
	
\end{thebibliography}

\end{document}